\documentclass[12pt]{iopart}
\usepackage{amsmath}
\usepackage{amssymb}
\usepackage{epsfig}

\begin{document}

\title{Dissipative quantum oscillator with two competing heat baths}
\author{Heiner Kohler\dag and Fernando Sols\ddag}

\begin{abstract}
We study the dissipative dynamics of a harmonic oscillator which couples
linearly through its position and its momentum to two independent heat baths
at the same temperature. We argue that this model describes a large
spin in a ferromagnet. We find
that some effects of the two heat baths partially cancel each other. This 
leads to unexpected features such as underdamped oscillations and long 
relaxation times in the strong coupling regime. Such a partial frustration 
of dissipation can be ascribed to the canonically conjugate character of 
position and momentum. We compare this model to the scenario where a single heat 
bath couples linearly to both the position and the momentum of the central 
oscillator. In that case less surprising behavior occurs for 
strong coupling. The dynamical evolution of the quantum purity for a 
single and a double wave packet is also investigated.
\end{abstract}


\address{\dag\
        Institut f\"{u}r theoretische Physik,
        Philosophenweg 19
        Universit\"at Heidelberg, Germany}
\address{\ddag\
         Departamento de F\'isica de Materiales ,
         Universidad Complutense de Madrid, E-28040 Madrid,
Spain}

\eads{\mailto{kohler@tphys.uni-heidelberg.de},\mailto{f.sols@fis.ucm.es}}

%\submitto{\NJP}

\maketitle

\section{Introduction}

\label{Intro}

The dissipative harmonic oscillator has for long attracted considerable
interest as a prototype of open quantum system. The early work of Magalinskii \cite{mag59}
and Ullersma \cite{ulle66a} focusing mainly on an Ohmic environment and the weak coupling
limit has been later extended in many different aspects \cite%
{mol70,dek77,tal81,schm82,grab83,grab84,lind84,ris85,haak85,grab88,unr89,har90} as, for instance,
to strong coupling \cite{haak85}, to non--Markovian noise \cite{lind84,grab88} or to nonlinear
coupling \cite{mol70}. Much activity was boosted by the work of Caldeira and Leggett \cite{cald83}
on the dissipative mechanics of a macroscopic quantum variable. Quantum decoherence has also been
addressed in Refs. \cite{hu92,zure93,rom97}. A comprehensive review can be found in the textbook by
Weiss \cite{weis99}\textit{.}

In the above context relatively little attention has been paid to effects
arising from the coupling to different system variables. The system variable
which couples to the heat bath is most often assumed to be the position
\textit{q}. In the sequel we refer to this model as the \textit{q--}%
oscillator. The choice of $q$ as the coupling variable was favoured in Ref.
\cite{cald83}. There it was argued that a complex dissipative environment
can be modelled by a bath of harmonic oscillators, with the couplings
parameters chosen to yield a Langevin equation for the $q$ variable in the
semiclassical limit. It was shown later \cite{ecke84} (see also Ref. \cite%
{kohl04}) that, in a superconducting Josepshon juntion, the oscillator bath
model can be derived microscopically from the coupling of the phase variable
to the quasiparticle bath, with the phase playing the role of position. At
that time the possibility of having a second bath coupled to the momentum $p$
variable was ruled out. However, in a superconducting weak link, the
electromagnetic field couples to the relative number variable, which in the
above scheme plays the role of momentum \cite{kohl04,zapa97}. If the
coupling to the phase is neglected, the semiclassical behavior of the
particle number is governed by an Abraham--Lorentz equation \cite%
{kohl04,jack75}. The coupling to the momentum has been also been considered
by Leggett \cite{legg84} and by Cuccoli \textit{et al.} \cite{cucc01}. A
systematic study of the combined role of the electromagnetic and
quasiparticle fields is presented in Ref. \cite{kohl04}. Thus Josephson
junctions provide a scenario where the distinction between the different
baths and their different coupling mechanisms becomes imperious.

Our study has also been motivated by the work of Castro Neto, Novais and
coworkers \cite{cast03}. To investigate the problem of a spin $\frac{1}{2}$
impurity in an antiferromagnetic environment, they studied a generalized
spin--boson problem using renormalization group techniques. The localized
spin couples with two components to the environment. They found a crossover
from incoherent to coherent transitions in the spin correlation functions in
certain parameter ranges. There the term \emph{quantum frustration} was
coined, which refers to the reduction of the effective interaction with the
environment due to the non--commutativity of the coupled operators. In this
article, we argue that a harmonic oscillator coupled through position and
momentum to two independent baths is a suitable representation of a large spin
impurity in a ferromagnetic system. We find a weaker cancellation of the
effects of the two baths. Since that situation corresponds to a large
(quasiclassical) spin, we have referred to this weaker form of frustration
as \emph{quasiclassical frustration }\cite{kohl05}.

Here we wish to investigate systematically the scenario where a particle couples through its
position and momentum to two baths, which are independent but have the same temperature. As a
paradigmatic system we choose a harmonic oscillator. This choice is motivated not only by
mathematical convenience. A more fundamental reason is that the symmetrical roles played by
position and momentum makes the harmonic oscillator the natural ground to study the difference and
the interplay (when both are present) between position and momentum coupling.

An explanatory comment is in order. As is well known, a coupling to $q$ can
be converted into a coupling to $p$ through a suitable canonical
transformation. However, the two representations bear an important
difference. Only in one of them is it possible to represent the system--bath
interaction as the coupling of the quantum variable to a system of \textit{%
otherwise independent} oscillators. A characteristic example is provided by
a particle of charge $e$ in an electromagnetic field, where a
\textquotedblleft velocity--coupling model" \cite{ford88} seems to apply.
Minimal coupling ($\mathbf{p\rightarrow p}-e\mathbf{A}/c$) generates not
only an interaction term $\mathbf{p\cdot A}$ but also a diamagnetic term $%
\propto \mathbf{A}^{2}$ which can be rightly interpreted as an interaction
between the effective oscillators. A unitary transformation $U=\exp (ie%
\mathbf{q\cdot A}/c),$ with $[q_{i},p_{j}]_{-}=i\delta _{ij}$, acting on $%
\mathbf{p}-(e/c)\mathbf{A}$, removes the coupling $\mathbf{p\cdot A}$. This
happens at the expense of generating a coupling $\propto \mathbf{q\cdot E}$
between the position and the electric field $\mathbf{E}$. Importantly, in
this new representation no quadratic field term is left, i.e. the charge
couples to a set of independent photons. Thus, in the precise language which
we propose here, a charged particle couples to the electromagnetic field
through its position $\mathbf{q}$.

In Sec.~\ref{Two independent baths} we study the model of an oscillator
coupled to two different baths. The general formalism is sketched in Sec.~%
\ref{General results}. In Sec.~\ref{ohm} we investigate the case where the
two independent Ohmic baths couple linearly to position and momentum. A
related feature is the possibility of inducing a crossover from overdamped
to underdamped oscillations by \textit{increasing} the damping coefficient.
Surprisingly, we obtain underdamped equilibrium oscillations for arbitrarily
strong coupling. Sec.~\ref{gencase} is devoted to the properties of spectral
functions under rather general combinations of double--bath systems.
Time dependent phenomena and implications for quantum decoherence are discussed
in Sec.~\ref{timeev} and Sec.~\ref{timedep} for the case of a double Ohmic bath. In Sec.~\ref{onebath} we
consider the situation where an oscillator couples to a single bath with the
most general form of linear coupling to both position and momentum. We find
that the most prominent feature of the double--bath model, namely, the
existence of underdamped oscillations for arbitrarily strong coupling,
disappears in the case of single--bath dissipation.

\section{Two independent baths}

\label{Two independent baths} The general form of a Hamiltonian
describing an oscillator coupled to two independent baths is \cite{kohl04}
\begin{equation}
H=\frac{\omega_p}{2}(p+\delta p)^2+\frac{\omega_q}{2}(q+\delta q)^2+\sum_{k}\omega _{k}a_{qk}^{\dagger
}a_{qk}+\sum_{k}\omega _{k}a_{pk}^{\dagger }a_{pk}\ ,  \label{mod1}
\end{equation}%
where $[q,p]_{-}=i$ and all operators are dimensionless and $\hbar =1$. The form of the Hamiltonian highlights the
symmetry between $q$ and $p$. The notion of mass is avoided by $m$ $=$ $1/\omega_p$. For the fluctuating pieces we assume
that they are linear in the bath variables and independent of $p$ and $q$,
\begin{eqnarray}
\delta q &=&i\sum_{k}\mu _{k}\left( a_{pk}^{\dagger
}-a_{pk}\right)  \notag \\
\delta p &=&i\sum_{k}\lambda _{k}\left( a_{qk}^{\dagger
}-a_{qk}\right) ~,  \label{mod2}
\end{eqnarray}%
% where $g_{q},g_{p}$ are sufficiently well--behaved functions. For general $%
% g_{q},g_{p}$ the two equations in Eq.~(\ref{mod2}) cannot be decoupled
% without creating all types of couplings between the two baths and one has to
% resort to a Taylor expansion of $g_{q},g_{p}$ in $\delta q$ and $\delta p$
% respectively. However, if one coupling function, say $g_{p}$, is equal to $1$%
% , we have $[\delta q,\delta p]_{-}=0$
After two unitary transformations $U_{p}=\exp (ip\delta q)$ and $U_{q}=\exp (iq\delta p)$ one
arrives at the Hamiltonian
% \begin{equation}
% \fl H=V_{q}(q)+\sum_{k}\omega _{k}\left\vert a_{qk}+\frac{\lambda _{k}}{%
% \omega _{k}}\int^{q}dq^{\prime }g_{q}(q^{\prime })\right\vert
% ^{2}+V_{p}(p)+\sum_{k}\omega _{k}\left\vert a_{pk}+\frac{\mu _{k}}{\omega
% _{k}}p\right\vert ^{2}\ ,  \label{mod4}
% \end{equation}%
% where the short-hand notation $a^{\dagger }a=|a|^{2}$ is used. Eq. (\ref%
% {mod4}) is a generalization of the Caldeira--Leggett Hamiltonian \cite%
% {cald83} to a situation where the system couples to two independent baths.
% Of course the momentum potential is almost always quadratic, $%
% V_{p}(p)=\omega _{p}p^{2}/2$. If we set also $V_{q}=\omega _{q}q^{2}/2$ and $%
% g_{q}=1$ we arrive at
\begin{equation}
H=\frac{\omega _{q}}{2}q^{2}+\sum_{k}\omega _{k}\left\vert a_{qk}+\frac{%
\lambda _{k}}{\omega _{k}}q\right\vert ^{2}+\frac{\omega _{p}}{2}%
p^{2}+\sum_{k}\omega _{k}\left\vert a_{pk}+\frac{\mu _{k}}{\omega _{k}}%
p\right\vert ^{2}\ , \label{mod5}
\end{equation}%
where the short-hand notation $a^{\dagger }a=|a|^{2}$ is used.
The model is a Caldeira--Leggett--type--of Hamiltonian \cite{cald83}.
It describes a harmonic oscillator with momentum $p$ and position $q$%
, each variable being coupled to a different oscillator bath. The frequency
of the central oscillator is $\omega _{0}=\left( \omega _{p}\omega
_{q}\right) ^{1/2}$. The baths are described by the spectral densities
\begin{equation}
J_{q}(\omega )\ =\ 2\sum |\lambda _{k}|^{2}\delta (\omega -\omega _{k})\
,\qquad J_{p}(\omega )\ =\ 2\sum |\mu _{k}|^{2}\delta (\omega -\omega _{k})\
.  \label{mod6}
\end{equation}%
Although one could have started directly with Eq.~(\ref{mod5}), we prefer to present the
Hamiltonian Eq.~(\ref{mod1}) as the starting point in order to provide a natural justification for
the renormalization terms $\sum_k|\mu_k|^2\omega_k^{-1}$ $p^2$ and
$\sum_k|\lambda_k|^2\omega_k^{-1}$ $q^2$ in Eq.~(\ref{mod5}) which otherwise would have to be
introduced {\em ad hoc}.

For low-lying excitations, and thus for low temperatures, the Hamiltonian (%
\ref{mod5}) becomes equivalent to that of a large spin $s$ in a
ferromagnetic environment. This can be seen as follows: The Hamiltonian for
a spin in a large magnetic field along the $z$ direction is
\begin{equation}
H^{\prime }\ =\ -\mu _{B}(S_{z}B_{z}+S_{x}\delta B_{x}+S_{y}\delta B_{y})+%
\mathbf{B}^{2},  \label{magnet1}
\end{equation}%
where the fluctuating terms model the low-lying bosonic (magnon) excitations
of the ferromagnet at the site of the large spin. The fluctuations in the $z$
direction are neglected since they are quadratic in $S_{x}$ and $S_{y}$ with
$S_{x}$, $S_{y}\ll S_{z}$. In other words $S_{z}$ is approximately a
constant of motion \cite{kit06}. For $s\rightarrow \infty $ the first term
in (\ref{magnet1}) is a harmonic oscillator $S_{z}=\hbar |a|^{2}-\hbar s$.
The action of $S_{i}$ on the eigenstates of this harmonic oscillator is
\begin{equation}
\begin{array}{ccccc}
S_{z}|n\rangle & = & (n-s)|n\rangle &  & \nonumber \\
S_{+}|n\rangle & = & \hbar \sqrt{2s-n}\sqrt{n+1}|n+1\rangle &  & \nonumber
\\
& \approx & \hbar \sqrt{2s}\sqrt{n+1}|n+1\rangle & \equiv & \hbar \sqrt{2s}%
a^{\dagger }\,|n\rangle \nonumber \\
S_{-}|n\rangle & = & \hbar \sqrt{2s-n+1}\sqrt{n}|n-1\rangle &  & \nonumber
\\
& \approx & \hbar \sqrt{2s}\sqrt{n}|n-1\rangle & \equiv & \hbar \sqrt{2s}%
a\,|n\rangle \ ,%
\end{array}
\label{holprim}
\end{equation}%
and $[S_{-},S_{+}]=2\hbar ^{2}s$. This amounts to keeping only the leading
order term in the Holstein--Primakoff transform of $S_{+}$, $S_{-}$ \cite%
{kit89}. The rescaled Hamiltonian $H^{\prime }/2s$ becomes formally
identical to Eq.~(\ref{mod5}) in the limit $s\rightarrow \infty $, with $%
\omega _{0}=$ $\mu _{B}B_{z}/2s$ and the fluctuating pieces scaling as $%
\delta B_{x}$ $=\sqrt{s}\sum \lambda _{k}(a_{qk}+a_{qk}^{\dagger })$ and $%
\delta B_{y}$ $=\sqrt{s}\sum \mu _{k}(a_{pk}+a_{pk}^{\dagger })$. Such large
spins have been observed in magnetic particles \cite{chud98}.

In the following we assume a power-law behavior at $\omega =0$ for both
spectral densities $J_{q}(\omega )$$=2\gamma _{q}\omega ^{\alpha _{q}}$$%
/(\omega _{\mathrm{ph}}^{\alpha _{q}-1}\pi )$ \ and $J_{p}(\omega )$$%
=2\gamma _{p}\omega ^{\alpha _{p}}$$/(\omega _{\mathrm{ph}}^{\alpha
_{p}-1}\pi )$. We introduce the \textquotedblleft phononic" frequency $%
\omega _{\mathrm{ph}}$ and the dimensionless coupling constants $\gamma _{n}$%
, with $n=$ $q,p$. Moreover we introduce a cutoff frequency $\Omega _{n}$
for both baths. Unless stated otherwise, we assume $\omega _{\mathrm{ph}}=$ $%
\Omega _{q}=\Omega _{p}\equiv \Omega $.

\subsection{General Results}

\label{General results} Elimination of the bath variables yields the
Heisenberg equations of motion for $q$ and $p$
\begin{eqnarray}
\dot{q}(t) &=&\omega _{p}p(t)+\int^{t}ds\,K_{p}(t-s)\dot{p}(s)+F_{p}(t)
\notag \\
-\dot{p}(t) &=&\omega _{q}q(t)+\int^{t}ds\,K_{q}(t-s)\dot{q}(s)+F_{q}(t).
\label{equations of motion}
\end{eqnarray}%
The response kernel is defined as%
\begin{equation}
K_{n}(t)\equiv \int_{0}^{\infty }\frac{J_{n}(\omega )}{\omega}
                     \cos (\omega t)d \omega \ ,\ n=q,p
\label{EM5}
\end{equation}%
and the force operator $F_{q}(t)=\sum \lambda _{k}a_{qk}\exp (-i\omega
_{k}t)+\mathrm{H.c}$., with $F_{p}(t)$ defined accordingly. In Fourier
space, Eq. (\ref{equations of motion}) reads%
\begin{align}
\left[ \widetilde{J}_{q}(\omega )-\omega _{q}\right] q+i\omega p&
=F_{q}(\omega )  \label{sist-Fourier} \\
-i\omega q+\left[ \widetilde{J}_{p}(\omega )-\omega _{p}\right] p&
=F_{p}(\omega )~,
\end{align}%
where $\widetilde{J}_{n}(\omega )$ is the symmetrized Riemann transform \cite%
{abr72}:
\begin{equation}
\widetilde{f}(\omega )\ \ =\ \omega ^{2}\mathcal{P}\int_{0}^{\infty }\frac{%
f(\omega ^{\prime })}{\omega ^{\prime }\left( {\omega ^{\prime }}^{2}-\omega
^{2}\right) }d\omega ^{\prime }-i\mathrm{sgn\,}(\omega )\frac{\pi }{2}%
f(|\omega |)~.  \label{tildetransformed}
\end{equation}%
The oscillation modes are given by the zeros of the function%
\begin{equation}
\chi ^{-1}(\omega )=\omega _{0}^{2}-\omega ^{2}-\omega _{q}\widetilde{J}%
_{p}(\omega )-\omega _{p}\widetilde{J}_{q}(\omega )+\widetilde{J}_{q}(\omega
)\widetilde{J}_{p}(\omega )  \label{ee13}
\end{equation}%
where $\chi (\omega )$ is the generalized susceptibility.

For the two baths at the same temperature
the symmetrized correlation functions for the position $C_{qq}^{(+)}(t)%
\equiv \frac{1}{2}\langle \lbrack q(t),q(0)]_{+}\rangle $ and the momentum $%
C_{pp}^{(+)}(t)\equiv \frac{1}{2}\langle \lbrack p(t),p(0)]_{+}\rangle $ are
obtained from $\chi (\omega )$ as follows
\begin{eqnarray}
C_{qq}^{(+)}(t) &=&\frac{1}{\pi }\int_{0}^{\infty }|\chi (\omega )|^{2}\cos
(\omega t)\coth (\beta \omega /2)  \notag \\
&&\qquad \qquad \mathrm{Im}\left( \widetilde{J}_{q}(\omega )\left\vert
\omega _{p}+\widetilde{J}_{p}(\omega )\right\vert ^{2}+\omega ^{2}\widetilde{%
J}_{p}(\omega )\right) d\omega \ .  \label{symmetrized autocorrelationfkt}
\end{eqnarray}%
The corresponding expression for $C_{pp}^{(+)}(t)$ is obtained by
interchanging $p\leftrightarrow q$ in Eq.~(\ref{symmetrized
autocorrelationfkt}). The antisymmetrized correlation functions for position
and momentum $C_{nn}^{(-)}(t)\equiv \langle \lbrack n(t),n(0)]_{-}\rangle $
(with $n=q,p$) are obtained from Eq.~(\ref{symmetrized autocorrelationfkt})
in the standard way \cite{weis99}, i.e. by substituting $\sin (\omega t)$
for $\cos (\omega t)\coth (\beta \omega /2)$.

We wish to emphasize that, within the dissipation model studied here, the 
equations of motion (\ref{equations of motion}) are symmetric in $q$ and $p$ for arbitrary coupling 
strength. Thus the situation is different from that of a $q$--oscillator when described within the 
rotating--wave approximation, which is known to yield a model symmetric in $q$ and $p$ but which holds only 
in the weak coupling limit \cite{gar00}.
%
%
%
%
%
%
%
%
%
%
%
%
%
%
%
%
%
%
%
%
%
%
%
%
%
%
%
%
%
%
%
%In general it is not possible to find analytic the exact
%expressions for the drift coefficients ${\bf f}_{nm}(t)$ and the
%diffusion coefficients ${\bf B}_{nm}(t)$ cannot be further
%simplified.
% In section \ref{coherent state} we discuss the time evolution of $W(q,p,t)$
% and $\mathcal{P}(t)$ for an initially coherent (Gaussian) state coupled to
% two Ohmic environments.

\subsection{Ohmic coupling}

\label{ohm}

\begin{figure}[b]
\begin{center}
\epsfig{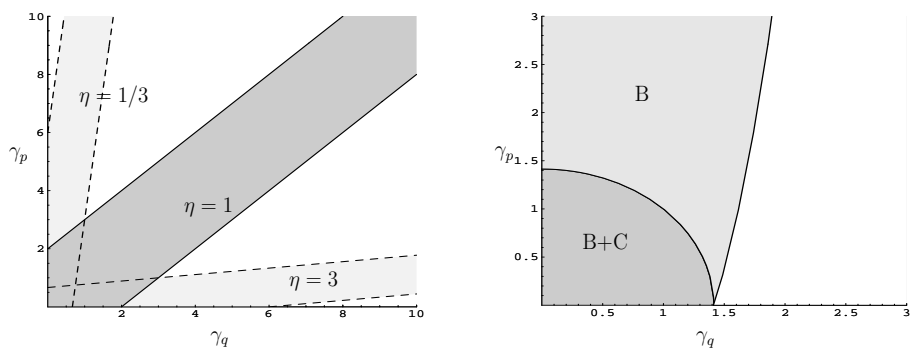}
\end{center}
\caption{Left: Stripes of underdamped oscillations in the $(\protect\gamma %
_{q},\protect\gamma _{p})$ plane for three different values of the parameter
$\protect\eta =\left( \protect\omega _{q}/\protect\omega _{p}\right) ^{1/2}$
$=m\protect\omega _{0}$, according to criterion A
[Eq. \ref{crit1}]. Right: Regions of underdamped oscillations with
criterions B [Eq.~(\protect\ref{othercrit})] and C [Eq.~(\protect
\ref{othercrit3})], for $\protect\eta =1$.}
% Left: Stripes of underdamped oscillations in the $(\protect\gamma %
% _{q},\protect\gamma _{p})$ plane for three different values of the parameter
% $\protect\eta =\left( \protect\omega _{q}/\protect\omega _{p}\right) ^{1/2}$
% $=M\protect\omega _{0}$. Right: Regions of underdamped oscillations with
% criterion B [Eq.~(\protect\ref{othercrit})] and criterion C [Eq.~(\protect
% \ref{othercrit3})], for $\protect\eta =1$}
\label{fig1}
\end{figure}

First we focus on the important special case that both spectral densities are Ohmic,
$\alpha_q$ $=$ $\alpha_p$ $=$ $1$ .
Then the real parts of $%
\widetilde{J}_{q}(\omega )$ and $\widetilde{J}_{p}(\omega )$ vanish to
lowest order in $\Omega ^{-1}$. The susceptibility has poles at the roots of
the quadratic polynomial \cite{tal81}
\begin{equation}
\omega _{0}^{2}-i(\omega _{q}\gamma _{p}+\omega _{p}\gamma _{q})\omega
-(1+\gamma _{q}\gamma _{p})\omega ^{2}~,  \label{charpol}
\end{equation}%
which are
\begin{equation}
\omega _{\pm }=\frac{\omega _{0}}{\left( 1+\gamma _{q}\gamma _{p}\right)
^{1/2}}\left( -i\kappa \pm \sqrt{1-\kappa ^{2}}\right) \ \equiv \ -i\tau
^{-1}\pm \zeta \ ,  \label{solu}
\end{equation}%
where
\begin{equation}
\kappa \equiv \frac{\omega _{p}\gamma _{q}+\omega _{q}\gamma _{p}}{2\omega
_{0}\left( 1+\gamma _{q}\gamma _{p}\right) ^{1/2}}.  \label{kappa}
\end{equation}%
The solutions of Eq.~(\ref{charpol}) are either purely imaginary or a pair
of complex conjugates depending on whether $\kappa $ is greater or smaller
than 1. Thus,
\begin{equation}
\kappa <1\qquad \mbox{Criterion A}\   \label{crit1}
\end{equation}%
is the commonly accepted criterion to distinguish between underdamped and
overdamped oscillations. The underdamped region lies in a stripe of width $%
\Delta =4\eta \left( 1+\eta ^{4}\right) ^{-1/2}$, with $\eta \equiv \left(
\omega _{q}/\omega _{p}\right) ^{1/2}$, limited by the graphs of the
functions
\begin{equation}
f(\gamma _{q})=\frac{\gamma _{q}}{\eta ^{2}}\pm \frac{2}{\eta }\ .
\label{solu1}
\end{equation}%
In Fig.~\ref{fig1} the stripes of underdamped oscillations, marked in the $%
(\gamma _{q},\gamma _{p})$ plane are plotted for three different values of $%
\eta $.
For large $\eta =m\omega _{0}$, corresponding to a large mass of the central
oscillator, the stripe of underdamped oscillations becomes 
smaller. In the limit $\eta \rightarrow \infty $, the introduction of an
infinesimal coupling to the momentum can induce a transition from overdamped
to underdamped oscillations. However the range of values of $\gamma _{p}$
allowing for underdamped oscillations also becomes increasingly small as $%
\propto \eta ^{-1}$.

The behavior of the system as a function of $\gamma _{p}$ for fixed $\gamma
_{q}$ is also interesting. We set $\eta =1$. The inverse damping time $\tau
^{-1}$ is a monotonously increasing function of $\gamma _{p}$ for $\gamma
_{q}<2$ and a monotonously decreasing function for $\gamma _{q}>2$.
Therefore an additional bath coupling to $p$ has opposite effects on the
damping time $\tau $ depending on whether the original $q$--oscillator
starts in the underdamped ($\gamma _{q}<2$) or in the overdamped ($\gamma
_{q}>2$) regime. In the first case, the additional bath always reduces the
damping time, leading to infinitely strong damping in the limit $\gamma
_{p}\rightarrow \infty $. However, in the regime $\gamma _{q}>2$ we can
drive the system from the overdamped into the underdamped regime by \emph{%
increasing} the coupling strength $\gamma _{p}$. What is more, we can take $%
\gamma _{p}=\gamma _{q}$ $=\gamma $, which corresponds to a completely
symmetric Hamiltonian in $p$ and $q$. At this point the system is \emph{%
always} in the underdamped regime and for large $\gamma $ the oscillator
frequency is close to its maximum. We find for the inverse damping time
\begin{equation}
\tau ^{-1}=\frac{\omega _{0}\gamma }{1+\gamma ^{2}}\ .  \label{tauspez}
\end{equation}%
Therefore we are led to the paradoxical situation that the higher the
friction coefficient $\gamma $ the larger $\tau $ becomes. In particular,
for $\gamma \rightarrow \infty $ one gets $\omega_0 \tau \rightarrow \infty $. 
This time dilatation is in itself a remarkable effect, since it 
contrasts with the behaviour of the $q$--oscillator, for which $\tau\to 0$ as $\gamma_q\to\infty$ 
[see Eq.~(\ref{solu})].

In the general, non--symmetric case, we note that for a given
finite $\gamma _{q}$ the damping time $\tau $ cannot be made to acquire arbitrary values
by tuning $\gamma_p$. Rather, it is bounded between $2\gamma _{q}/\omega _{0}$ and  
$2/\gamma _{q}\omega _{0}$, which is higher depending on $\gamma_q \gtrless 1$.

%for $\gamma _{q}>1$ for $\gamma _{q}<1$.
 %from abov by $2\gamma _{q}/\omega _{0}$ for $%
%\gamma _{q}>1$ and by $2/\gamma _{q}\omega _{0}$ for $\gamma _{q}<1$.
However, from these striking observations one should not conclude that in the
limit of infinite $\gamma $ the central oscillator recovers the dynamics of
a free oscillator. It rather leads to a dilatation of all time scales. For
instance, the renormalized oscillator frequency $\zeta $ vanishes even
faster than the inverse damping time, since $\zeta =\omega _{0}/(1+\gamma
^{2})$. An analysis of the correlator in Eq.~(\ref{symmetrized
autocorrelationfkt}) shows that for $t\ll \tau $ the particle behaves as a
free ballistic (though very slow) particle, $C_{qq}^{(+)}(t)\propto t/\tau $.
% . From the enhancement of the damping time, it also follows an enhancement
% of the decoherence time through the relation $\tau _{\varphi }\propto \beta
% \tau $ which holds generally for open quantum systems at high enough
% temperatures \cite{weis99}.

Instead of using criterion A, we may focus on $D_{q}(\omega )\equiv \mathrm{%
Im}C_{qq}^{(-)}(\omega )/\omega $. As a spectral function, $\mathrm{Im}%
C_{qq}^{(-)}(\omega )$ is independent of temperature. 
%In the classical limit it is always zero.
Of special interest is the slope of $D_{q}(\omega )$
near $\omega =0$. Since $\lim_{\omega\rightarrow \infty }D_{q}(\omega )=0$,
the condition for the existence for $%
D_{q}(\omega )$ displaying a maximum can be written as%
\begin{equation}
D_{q}^{\prime }(0)>0\qquad \mbox{Criterion B}~,  \label{othercrit}
\end{equation}%
which may be viewed as an indicator of underdamped oscillations or,
equivalently, coherent transitions. It is often employed for the spin--boson model \cite{legg84,cast03}.
For Ohmic damping, one finds
\begin{equation}
D_{q}(\omega )=\frac{\gamma _{q}\omega _{p}^{2}+\gamma _{p}(1+\gamma
_{q}\gamma _{p})\omega ^{2}}{[(1+\gamma _{q}\gamma _{p})\omega ^{2}-\omega
_{0}^{2}]^{2}+(\gamma _{q}\omega _{p}+\gamma _{p}\omega _{q})^{2}\omega ^{2}}%
~,  \label{crit2}
\end{equation}%
and the critical curve for $\gamma _{p}$ is thus given by the relation $%
\gamma _{p}^{\text{crit}}=$ $\gamma _{q}$ $(\gamma _{q}^{2}/\eta ^{2}-2)$ $%
/\eta ^{2}$, so that criterion B is satisfied for $\gamma _{p}>\gamma _{p}^{%
\text{crit}}$.

\begin{figure}[b]
\par
\begin{center}
\epsfig{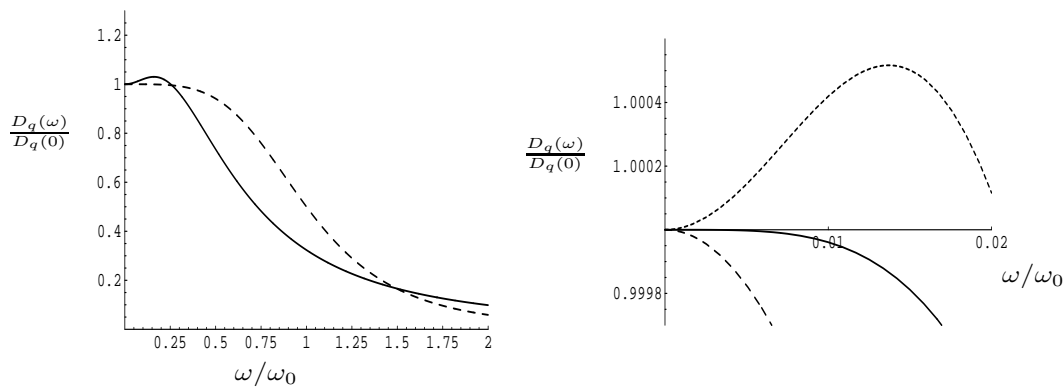}
\caption{The spectral function $D_{q}(\protect\omega )$ for different
coupling strengths $\protect\gamma _{p}$, $\protect\gamma _{q}$. On the
left, the marginal case of the $q$--oscillator, at $(\protect\gamma _{q},%
\protect\gamma _{p})=(\protect\sqrt{2},0)$, (\dashed), develops a maximum
for $(\protect\gamma _{q},\protect\gamma _{p})=(\protect\sqrt{2},\protect%
\sqrt{2})$, (\full). On the right, for a value of $\protect\gamma _{q}=3$
(overdamped regime), when the values of $\protect\gamma _{p}$ are $0$ (
\dashed), $21$ (\full) (marginal case), and $40$ (\dotted).}\label{fig2}
\end{center}
\par
% The normalized Fourier transform $D_{q}(\protect\omega )/\protect%
% \omega $ of the autocorrelation function for different coupling strength $%
% \protect\gamma _{p}$, $\protect\gamma _{q}$. On the left, the marginal case
% of the $q$--oscillator, at $(\protect\gamma _{q},\protect\gamma _{p})=(%
% \protect\sqrt{2},0)$, (\dashed), develops a maximum for $(\protect\gamma %
% _{q},\protect\gamma _{p})=(\protect\sqrt{2},\protect\sqrt{2})$, (\full). On
% the right for a value of $\protect\gamma _{q}=3$ in the overdamped regime.
% The values of $\protect\gamma _{p}$ are $0$ (\dashed), $21$ (\full)
% (marginal case), and $40$ (\dotted).
\end{figure}

Both being based on exact expressions, we observe that criterion B differs
substantially from criterion A, see Fig.~\ref{fig1}. Surprisingly, even for
the $q$--oscillator there is a region $\sqrt{2}<\gamma _{q}<2$ where
criterion A and criterion B are different. In Fig. \ref{fig2} $D_{q}(\omega
) $ is plotted for different coupling strengths $\gamma _{q}$, $\gamma _{p}$%
.
For large couplings $\gamma _{p}$, we observe that, although we are
(according to criterion B) in the underdamped region, the maxima of the
curves are not very pronounced. This is related to the fact that the
renormalized oscillator frequency $\zeta $ vanishes for large $\gamma _{p}$
faster than the inverse damping time $\tau ^{-1}$.

The foregoing discussion suggests the introduction of a third, more
restrictive criterion for underdamped oscillations, namely \cite{kohl05},
\begin{equation}
\zeta >\tau ^{-1}\qquad \mbox{Criterion C.}  \label{othercrit3}
\end{equation}%
According to this condition, the region of underdamped oscillations is
convex, i.e. by increasing the coupling strength no transition from
overdamped to underdamped oscillations can be realized.
\begin{table}[tbp]
\centering
%EndExpansion%
\begin{tabular}{cc|c|c|c}
&  & symmetric & only $\gamma_{q}$ & only $\gamma_{p}$ \\ \hline
$\kappa<1$ & (A) & \textit{always} & $\gamma_{q}<2\eta$ & $%
\gamma_{p}<2\eta^{-1}$ \\ \hline
$D_{q}(\omega)$ has peak & (B) & $\gamma<\sqrt{3}$ & $\gamma_{q}<\sqrt{2}%
\eta $ & \textit{always} \\ \hline
$|\mathrm{Im}\omega_{\pm}|<|\mathrm{Re}\omega_{\pm}|$ & (C) & $\gamma<1$ & $%
\gamma_{q}<\sqrt{2}\eta$ & $\gamma_{p}<\sqrt{2}\eta^{-1}$%
\end{tabular}%
\caption{Condition for coherent dynamics according to three different
criteria (left column, labeled A, B, C), for three particular cases (upper
row). The symmetric limit includes the assumption $\protect\eta =1$. General
case is given in main text. }
\label{TableKey-5}
\end{table}
Table I summarizes the different criteria for coherent, underdamped dynamics
as they are applied to three prototypical cases.

For high temperatures ($k_{B}T\gg \omega _{0}$) the integral for the
coordinate autocorrelation function Eq.~(\ref{symmetrized autocorrelationfkt}%
) can easily be solved yielding the classical solutions of the Heisenberg
equations of motion. For $\kappa <1$ they read
\begin{eqnarray}
\fl C_{qq}^{(+)}(t)/\langle q^{2}\rangle \ =\ \left\{ \cos \left( \frac{%
\omega _{0}\sqrt{1-\kappa ^{2}}}{\sqrt{1+\gamma _{q}\gamma _{p}}}t\right) +%
\frac{\omega _{p}\gamma _{q}-\omega _{q}\gamma _{p}}{2\omega _{0}\sqrt{%
\left( 1-\kappa ^{2}\right) \left( 1+\gamma _{q}\gamma _{p}\right) }}\sin
\left( \frac{\omega _{0}\sqrt{1-\kappa ^{2}}}{\sqrt{1+\gamma _{q}\gamma _{p}}%
}t\right) \right\} &&  \notag  \label{classical} \\
\qquad \qquad \qquad \exp \left( \frac{-\omega _{0}\kappa |t|}{\sqrt{%
1+\gamma _{q}\gamma _{p}}}\right) \ . &&
\label{hightemp}
\end{eqnarray}%
For $\kappa >1$, Eq.~(\ref{classical}) describes the corresponding
non-oscillating solution. Analytical expressions for $C_{qq}^{(+)}(t)$ are
also available for $T=0$. In this case the integral in Eq.~(\ref{symmetrized
autocorrelationfkt}) can be expressed in terms of exponential integrals \cite%
{weis99}. We do not show the result here and limit ourselves to note that 
for long times only the lowest order term, i.~e. the term linear in $\omega $%
, contributes in the numerator of the r.h.s. of Eq.~(\ref{symmetrized
autocorrelationfkt}). Thus the long time behaviour is the same as for the $q$%
--oscillator, namely, we have
\begin{equation}
\label{CTzero} C_{qq}^{(+)}(t)= -\frac{\omega_p^2\gamma_p}{2\pi}\frac{1}{\omega_0t^{2}} + {\cal
O}(t^{-3})\ , \quad {\rm for}\quad t\rightarrow \infty .
\end{equation}
For the classical solution Eq.~(\ref{hightemp}) criterion A [Eq.~(\ref{crit1})] is the only natural
criterion to differentiate between underdamped and overdamped oscillations, since it distinguishes
solutions with infinitely many zeros from solutions with no zeros at all.  For zero temperature the
number of zeros of $C_{qq}^{(+)}$ is always finite and criterion A is less informative.

The zero temperature mean squares $\langle q^{2}\rangle $, $\langle
p^{2}\rangle $ can be calculated exactly. They are given by
\begin{eqnarray}
\langle q^{2}\rangle &=&\frac{f(\kappa )}{4\kappa (1+\gamma _{q}\gamma _{p})}%
\left[ \frac{\gamma _{q}}{\eta ^{2}}+\gamma _{p}\left( 1-2\kappa ^{2}\right) %
\right]  \notag  \label{meanq} \\
&&+\frac{\gamma _{p}}{\pi }\left( 1+\gamma _{p}\gamma _{q}\right) \ln \frac{%
\Omega _{p}\sqrt{1+\gamma _{q}\gamma _{p}}}{\omega _{0}}\ +\mathcal{O}%
(\Omega _{p}^{-1},\Omega _{q}^{-1}),  \notag \\
f(\kappa ) &=&\frac{1}{\pi \sqrt{\kappa ^{2}-1}}\ln \frac{\kappa +\sqrt{%
\kappa ^{2}-1}}{\kappa -\sqrt{\kappa ^{2}-1}}\ .
\end{eqnarray}%
The corresponding expressions for $\langle p^{2}\rangle $ are obtained by
interchanging $p\leftrightarrow q$ in Eqs.~(\ref{meanq}) and (\ref{kappa}).
For small $\gamma _{q},\gamma _{p}$ we have $f(\kappa )=1-2\kappa /\pi +%
\mathcal{O}(\kappa ^{2})$ and the position mean square becomes
\begin{equation}
\langle q^{2}\rangle =\frac{1}{2\eta }-\frac{\gamma _{q}}{2\eta ^{2}}+\gamma
_{p}\left( \ln \frac{\Omega _{p}}{\omega _{0}}-\frac{1}{2}\right) +\mathcal{O%
}(\gamma _{q},\gamma _{p})\ ,  \label{meanqper}
\end{equation}%
and, correspondingly,%
\begin{equation}
\langle p^{2}\rangle =\frac{\eta }{2}-\frac{\eta ^{2}}{2}\gamma _{p}+\gamma
_{q}\left( \ln \frac{\Omega _{q}}{\omega _{0}}-\frac{1}{2}\right) +\mathcal{O%
}(\gamma _{q},\gamma _{p})\ ,
\end{equation}%
For $\gamma _{q}\neq 0$, $\gamma _{p}=0$ we recover the results for the $q$%
--oscillator, with the characteristic logarithmic dependency of $\langle
p^{2}\rangle $ on the cutoff \cite{weis99}. In the general case, the
Heisenberg uncertainty product diverges as $\langle p^{2}\rangle \langle
q^{2}\rangle \propto \ln \Omega _{p}\ln \Omega _{q}$. Thus the reduced
density matrix at equilibrium becomes approximately the identity, its
off--diagonal elements being essentially zero in both the position and the
momentum representation.

\subsection{Other spectral Densities}

\label{gencase}

For general spectral densities $\alpha _{q}$ and $\alpha _{p}$ are arbitrary
positive real numbers. While we always have
\begin{equation}
\mathrm{Im}\widetilde{J}_{n}(\omega )=\frac{\pi }{2}\mathrm{sgn\,}(\omega
)J_{n}(|\omega |)~,  \label{gen1}
\end{equation}%
we find
\begin{eqnarray}
\mathrm{Re}\widetilde{J}_{n}(\omega ) &=& \left\{%
\begin{matrix}
0 \ , & \ \text{if \ }\alpha _{n}<2 , \cr \frac{\gamma _{n}\omega ^{2}}{\pi
\omega _{\mathrm{ph}}}\, \ln \frac{\Omega_{n}^{2}}{\omega ^{2}}\ , & \ ~%
\text{if \ }\alpha _{n}=2\ , \cr \frac{2\gamma _{n}\Omega _{n}^{\alpha
_{n}-2}}{\pi (\alpha _{n}-2)\omega _{\mathrm{ph}}^{\alpha _{n}-1}}\,\omega
^{2}\ , & ~\text{if \ }\alpha _{n}>2\ , & \quad \ n=q,p%
\end{matrix}%
\right. .  \label{ReJn}
\end{eqnarray}

%In the subsequent discussion we focus on the zero temperature equilibrium
%position mean square given by Eq.~(\ref{symmetrized autocorrelationfkt}).
First, we briefly recall the behavior of $\langle q^{2}\rangle $ and $%
\langle p^{2}\rangle $ as functions of $\alpha _{q}$ and $\gamma _{q}$ in
the case of the $q$--oscillator. For $\langle q^{2}\rangle $ it may be
summarized by the formula \cite{weis99}
\begin{equation}
\frac{d\langle q^{2}\rangle }{d\gamma _{q}}\quad \left\{
\begin{array}{ccc}
< & 0 & ~\text{for}~~\alpha _{q}<2 \\
\propto & -\frac{1}{\Omega } & ~\text{for ~}\alpha _{q}\geq 2%
\end{array}%
\right. \ .  \label{pcmeansq}
\end{equation}%
A measurement of the position, i.e. a diagonalization of the density matrix
in the position basis takes place only for $\alpha _{q}<2$. The behaviour of
$\langle p^{2}\rangle $ is opposed to that of $\langle q^{2}\rangle $ in
such a way that the product $\langle q^{2}\rangle \langle p^{2}\rangle $ is
always $\geq 1/4$, as required by the Heisenberg relation. For $\langle
p^{2}\rangle $ we may write
\begin{equation}
\frac{d\langle p^{2}\rangle }{d\gamma _{q}}\quad \left\{
\begin{array}{ccc}
\propto & \ln \left( \frac{\Omega }{\omega _{0}}\right) & ~\text{for}%
~~\alpha _{q}\leq 2 \\
> & 0 & ~\text{for ~}\alpha _{q}> 2%
\end{array}%
\right. \ .  \label{pcmommeansq}
\end{equation}

In the forthcoming discussion we focus on $\langle q^{2}\rangle $. We have a
contribution from $J_{q}(\omega )$, which reduces $\langle q^{2}\rangle $,
and another contribution from $J_{p}(\omega )$ enhancing $\langle
q^{2}\rangle $ [see Eq. (\ref{meanqper})]. From the results for the $q$%
--oscillator we expect that the $J_{p}(\omega )$ contribution should always
dominate for $\alpha _{p}<2$ and the $J_{q}(\omega )$ contribution should be
negligible for $\alpha _{q}\geq 2$. The opposite regime $\alpha _{q}<2$ and $%
\alpha _{p}>2$ is more interesting (the case $\alpha _{q}=1$ and
$\alpha _{p}=3$ has been studied in Ref. \cite{kohl04}) . There we
have a sum of two terms of order $1$ in $\Omega ^{-1}$ with opposite
sign. In the following we focus on that case.

We assume the susceptibility $\chi (\omega )$ to be an algebraic function.
This entails rational exponents $\alpha _{n}=\beta _{n}/m$ with integer $%
\beta _{n}$ and $m$. The integrand in Eq.~(\ref{symmetrized
autocorrelationfkt}) has $r\equiv \max (\beta _{q}+\beta _{p},2m)$ poles on
the $m$ Riemann sheets which split into two classes. For $\alpha _{q},\alpha
_{p}>1$ there are $2m$ poles close to the oscillator frequency $\pm \omega
_{0}$ on each sheet. In addition there are $r_{\Omega }\equiv \max (\beta
_{q}+\beta _{p}-2m,0)$ poles of the order of $i\Omega $. If either $\alpha
_{q}$ or $\alpha _{p}$ becomes equal or smaller one, the $2m$ poles
detach from $\pm \omega _{0}$ moving into the complex plane. The
equilibrium mean square can be written as a sum of the contributions from
the two types of poles.
\begin{equation}
\langle q^{2}\rangle =\mathcal{C}_{\omega _{0}}+\mathcal{C}_{\Omega }\ .
\label{expression for general spectralfunctions}
\end{equation}%
For the second term we find ($\gamma _{q}$, $\gamma _{p}$ small)
\begin{equation}
\mathcal{C}_{\Omega }=\frac{\gamma _{p}}{\pi }\frac{1}{\alpha _{p}-1}\Theta
(\alpha _{p}-1)+\mathcal{O}(\gamma _{p},\gamma _{q})\ .  \label{highcont}
\end{equation}%
For a sketch of the derivation of Eq.~(\ref{highcont}) see \ref{appA}.
Therefore, the high energy poles yield, to first order in $\gamma _{p}$, a
finite contribution to the position mean square. This contribution is
positive for superohmic coupling to the momentum but vanishes for Ohmic and
subohmic coupling. It has a singularity for $\alpha _{p}=1$ which marks the
transition to the logarithmic dependence on the cutoff frequency, see Eq.~(%
\ref{meanq}). We note that the right hand side of Eq.~(\ref{highcont}) does
not depend on the oscillator frequency $\omega _{0}$, i.~e. it does not
depend on the properties of the system itself.
\begin{figure}[tbp]

\label{fig4}
\begin{center}
\epsfig{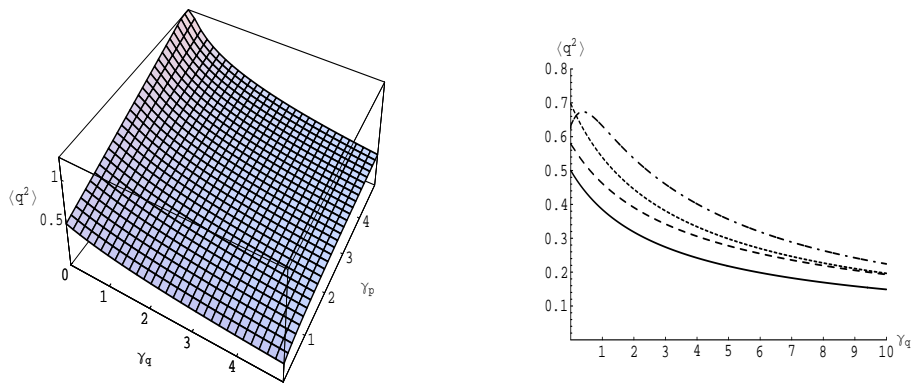}
\end{center}
\caption{Left: Zero temperature mean square $\langle q^{2}\rangle $ as a
function of $\protect\gamma _{q}$ and $\protect\gamma _{p}$ for $\protect%
\alpha _{q}=1$ (Ohmic) and $\protect\alpha _{p}=3$ (superohmic). Right: Zero
temperature mean square $\langle q^{2}\rangle $ as a function of $\protect%
\gamma _{q} $ for $\protect\alpha _{q}=1$ , $\protect\gamma_p=1$ and
different spectral exponents $\protect\alpha _{p}=3$ (\dashed), $\protect%
\alpha _{p}=2.2$ (\chain) and $\protect\alpha _{p}=1.8$ (\dotted). The full
line corresponds to the $q$--oscillator $\protect\gamma_p=0$. The cutoff
frequency is $\Omega /\protect\omega _{0}=200$ in both figures. The case $%
\protect\alpha _{p}=2$ is avoided because of its singular character [see Eq.
(\protect\ref{ReJn})].}
%  {Left: Zero temperature mean square $\langle q^{2}\rangle $ as a
% function of $\protect\gamma _{q}$ and $\protect\gamma _{p}$ for $\protect%
% \alpha _{q}=1$ (Ohmic) and $\protect\alpha _{p}=3$ (superohmic). Right: Zero
% temperature mean square $\langle q^{2}\rangle $ as a function of $\protect%
% \gamma _{q}$ for $\protect\alpha _{q}=1$ and different different spectral
% exponents $\protect\alpha _{p}=3$ (\dashed), $\protect\alpha _{p}=2.2$ (
% \chain) and $\protect\alpha _{p}=1.8$ (\dotted). The full line corresponds
% to the $q$--oscillator $\protect\gamma _{p}=0$. The cutoff frequency is $%
% \Omega /\protect\omega _{0}=200$ in both figures. The case $\protect\alpha %
% _{p}=2$ is avoided because of its singular character [see Eq. (\protect\ref%
% {ReJn})].}
\end{figure}
The contribution from the poles close to $\omega _{0}$, $\mathcal{C}_{\omega
_{0}}$, can in principle be calculated in a similar way as $\mathcal{C}%
_{\Omega }$ for arbitrary exponents $\alpha _{q}$, $\alpha _{p}$; however,
the calculations become increasingly messy. A general treatment is also
hampered by the wide variety of casuistic behavior, with different regimes
defined by the conditions $\alpha _{n}\gtrless 1$, $\alpha _{n}\gtrless 2$
and $\alpha _{q}+\alpha _{p}\gtrless 2$, $\alpha _{q}+\alpha _{p}\gtrless 4$.

We focus on the specific case $\alpha _{q}\leq 1$ and $\alpha _{p}>2$, which
is the most relevant, since most baths occuring in nature are either Ohmic
(Markovian approximation, electron gas \cite{sols87}) or superohmic (photon
or acoustic phonon baths). In that case, the low energy poles are
essentially determined by the lower spectral exponent (here, $\alpha _{q}$).
This reflects the general property of dissipative quantum systems that an
environment becomes increasingly efficient with decreasing spectral exponent
$\alpha _{n}$. To leading order in $\gamma _{q}$, we find
\begin{equation}
C_{\omega _{0}}=\frac{1}{2}-\frac{\gamma _{q}}{\pi \alpha _{q}^{2}}\left[
\frac{1}{2}+3\alpha _{q}\left( \alpha _{q}-1\right) \right] +\mathcal{O}%
(\gamma _{q}^{2})\ .  \label{lowcont}
\end{equation}%
If we compare $\mathcal{C}_{\omega _{0}}$ with Eq.~(\ref{highcont}), we
observe that the two contributions have opposite sign. In particular, the
contributions of $C_{\omega _{0}}$ and $C_{\Omega }$ \textit{cancel each
other} to first order in $\gamma _{q}$ and $\gamma _{p}$ provided that
\begin{equation}
\gamma _{p}=\frac{\alpha _{p}-1}{\alpha _{q}^{2}}\left[ \frac{1}{2}+3\alpha
_{q}\left( \alpha _{q}-1\right) \right] \gamma _{q}\ .  \label{equalgam}
\end{equation}%
In Fig.~\ref{fig4} $\langle q^{2}\rangle $ is plotted against the coupling
strengths $\gamma _{p}$ and $\gamma _{q}$ for $\alpha _{q}=1$ and $\alpha
_{p}=3$. $\langle q^{2}\rangle $ is monotonously increasing as a function of
$\gamma _{p}$. For $\gamma _{p}=0$ it is a monotonously decreasing function
of $\gamma _{q}$ . For the parameters chosen in \Fref{fig4}, \Eref{equalgam}
holds on the diagonal $\gamma _{q}$ $=\gamma _{p}$. There $\langle
q^{2}\rangle $ remains close to its unperturbed value $1/2$ if $\gamma
_{q},\gamma _{p}\lesssim 1$.

Finally, on the right hand side of \Fref{fig4} the position mean square $%
\langle q^{2}\rangle $ is plotted as a function of $\gamma _{q}$ for
different spectral exponents $\alpha _{p}$. For small $\gamma _{q}$ the
enhancement of $\langle q^{2}\rangle $ due to the coupling to the momentum
is larger for smaller values of the spectral exponent $\alpha _{p}$. This is
the expected behaviour. This behavior, however, is inverted when $\gamma
_{q} $ increases. Then the relative enhancement of $\langle q^{2}\rangle $
as compared with the $q$--oscillator is bigger for higher spectral exponent $%
\alpha _{p}$. A crossing occurs always for some value of the coupling
constant $\gamma _{q}$, although some times this can be very large.

\section{Time evolution}
\label{timeev}

To study the nonequilibrium properties and in particular the loss of
coherence for an initially pure state, the calculation of $C_{qq}^{(+)}(t)$
is not sufficient. Instead, one should perform a full nonequilibrium
calculation with inclusion of specific initial conditions by means of a
Laplace transformation. The position-momentum symmetry of the oscillator
Hamiltonian suggests the use of the reduced Wigner--function $W(q,p,t)$
instead of the reduced density matrix. It was shown in Ref.~\cite{kar97}
for the $q$ oscillator that the time evolution of $W(q,p,t)$ can only for certain initial
conditions be described by an exact master equation. Here we consider decoupled initial
conditions $W_{0}=W_{S}W_{B}^{(q)}W_{B}^{(p)}$,  which fall into this class. In this case
we are able to derive not only an exact Master equation for the reduced Wigner function but a solution in terms of a
two--fold convolution integral.

Generally $W(q,p,t)$ can be expressed
as
\begin{equation}
W(q,p,t)=\langle \delta (q-q(t))\delta (p-p(t))\rangle _{0}\ ,
\label{reduced Wgnerfunktion}
\end{equation}%
where the bracket denotes the average over initial conditions \cite{haak85}
\begin{equation}
\fl\langle \ldots \rangle _{0}\equiv \int dq^{\prime }dp^{\prime
}\prod_{k,k^{\prime }}da_{qk}da_{pk^{\prime }}da_{qk}^{\ast }da_{pk^{\prime
}}^{\ast }(\ldots )W_{S}(q^{\prime },p^{\prime },\{\mathbf{a}_{q},\mathbf{a}%
_{p},\mathbf{a}_{q}^{\ast },\mathbf{a}_{p}^{\ast }\})\ ,
\end{equation}%
and $q(t),p(t)$ are the classical solutions of the equations
of motion Eq.~({\ref{equations of motion}). The Wigner function of the thermalized baths
is given by
\begin{equation}
W_{B}^{(n)}=\prod_{k}\frac{i}{\pi \coth (\beta \omega _{k}/2)}\exp \left( -%
\frac{2a_{nk}^{\ast }a_{nk}}{\coth (\beta \omega _{k}/2)}\right) \ ,
\label{thermal}
\end{equation}%
the system itself being in an arbitrary pure state characterized by $W_{S}$.
In this case one can express Eq.~(\ref{reduced Wgnerfunktion}) as a twofold
convolution integral
\begin{eqnarray}
\fl W(q,p,t) &=&\int \frac{dk}{2\pi }\frac{dk^{\prime }}{2\pi }%
e^{ikp+ik^{\prime }q}\mathcal{W}_{S}(k\phi _{0}(t)+k^{\prime }\phi
_{p}(t),k^{\prime }\phi _{0}(t)-k\phi _{q}(t))  \notag  \label{wigsolution}
\\
&&\lo\qquad \qquad \exp \left( -k^{2}X_{pq}(t)-{k^{\prime }}%
^{2}X_{p0}(t)+2kk^{\prime }Y_{pq}(t)\right)  \notag \\
&&\lo\qquad \qquad \exp \left( -{k^{\prime }}%
^{2}X_{qp}(t)-k^{2}X_{q0}(t)-2kk^{\prime }Y_{qp}(t)\right) \ ,
\end{eqnarray}%
where $\mathcal{W}_{S}$ is the Fourier transform of $W_{S}$ in both
arguments. We have introduced the auxiliary functions
\begin{eqnarray}
\phi _{n}(t) &\equiv &\frac{1}{2\pi i}\int d\omega \chi (\omega )\left(
\omega _{n}+\widetilde{J}_{n}(\omega )\right) e^{i\omega t}~,~~~~n=q,p
\notag  \label{definition of auxiliary functions phi_0,phi_p} \\
\phi _{0}(t) &\equiv &\frac{1}{2\pi i}\int d\omega \omega \chi (\omega
)e^{i\omega t}\ ,
\end{eqnarray}%
and the temperature-dependent quantities
\begin{eqnarray}
X_{pq}(t) &=&\frac{1}{2}\int_{0}^{\infty }\,d\omega J_{p}(\omega )\coth
(\beta \omega /2)\left\vert \int_{0}^{t}\,dt\phi _{q}(t-t^{\prime
})e^{i\omega t}\right\vert ^{2}  \notag \\
X_{p0}(t) &=&\frac{1}{2}\int_{0}^{\infty }\,d\omega J_{p}(\omega )\coth
(\beta \omega /2)\left\vert \int_{0}^{t}\,dt\phi _{0}(t-t^{\prime
})e^{i\omega t}\right\vert ^{2}  \notag \\
Y_{pq}(t) &=&\frac{1}{2}\int_{0}^{\infty }\,d\omega J_{p}(\omega )\coth
(\beta \omega /2)  \notag \\
&&\mathrm{Re}\left[ \int_{0}^{t}\,dt\phi _{0}(t-t^{\prime })e^{i\omega
t}\int_{0}^{t}\,dt\phi _{q}(t-t^{\prime })e^{-i\omega t}\right] \ .
\label{tempdep}
\end{eqnarray}%
The functions $X_{q0}$, $X_{qp}$ and $Y_{qp}$ are defined accordingly by
interchanging $p$ and $q$. For a better understanding of the forthcoming
discussion in Sec.~\ref{timedep} we notice that the functions defined
in Eq.~(\ref{definition of auxiliary functions phi_0,phi_p}) are the
time--dependent coefficients of the solutions $q(t)$, $p(t)$ of the initial
value problem Eq.~(\ref{equations of motion}). For instance
\begin{equation}
\fl\quad q(t)=\phi _{0}(t)q(0)+\phi _{p}(t)p(0)+\int_{0}^{t}\phi
_{0}(t-t^{\prime })F_p(t^{\prime })dt^{\prime }+\int_{0}^{t}\phi
_{p}(t-t^{\prime })F_q(t^{\prime })dt^{\prime }\ .  \label{laplace}
\end{equation}%
This necessarily requires $\phi _{0}(0)=1$ and $\phi _{n}(0)=0$. Finally, we
note that $W$ fulfills a Fokker--Planck type of equation \cite{weis99},
\begin{equation}
\dot{W}(q,p,t)=\nabla \left[ \mathbf{g}(q,p,t)+\mathbf{B}(t)\nabla \right]
W(q,p,t)\ ,  \label{Fokker--Planck equation}
\end{equation}%
where }$\nabla \equiv (\partial /\partial q,\partial /\partial p)$, {with $%
\mathbf{g}=(g_{q},g_{p})$ the phase--space drift term,
\begin{eqnarray}
g_{n}(q,p,t) &=&-\sum_{m=q,p}G_{nm}(t)m~, \\
\mathbf{G}(t) &\equiv &\frac{\left(
\begin{array}{cc}
{\dot{\phi}}_{0}\phi _{0}+{\dot{\phi}}_{p}\phi _{q} & {\dot{\phi}}_{p}\phi
_{0}-{\dot{\phi}}_{0}\phi _{p} \\
{\dot{\phi}}_{0}\phi _{q}-{\dot{\phi}}_{q}\phi _{0} & {\dot{\phi}}_{0}\phi
_{0}+{\dot{\phi}}_{q}\phi _{p}%
\end{array}%
\right) }{\phi _{0}^{2}+\phi _{q}\phi _{p}}~,  \label{driftterm}
\end{eqnarray}%
and with (the $2\times 2$ matrix) $\mathbf{B}(t)$ the state--independent
phase--space diffusion term,
\begin{eqnarray}
B_{pp}(t) &=&\dot{X}_{pq}+\dot{X}%
_{q0}-2F_{pp}(X_{pq}+X_{q0})+2F_{pq}(Y_{pq}-Y_{qp})  \label{diffusionterm} \\
B_{qq}(t) &=&\dot{X}_{qp}+\dot{X}%
_{p0}-2F_{qq}(X_{qp}+X_{p0})-2F_{qp}(Y_{qp}-Y_{pq})  \notag \\
B_{pq}(t) &=&\dot{Y}_{qp}-\dot{Y}_{pq}-(F_{pp}+F_{qq})(Y_{qp}-Y_{pq})  \notag
\\
&&-F_{pq}(X_{p0}+X_{qp})-F_{qp}(X_{pq}+X_{q0}).
\end{eqnarray}%
}$B_{qp}(t)$ is obtained from $B_{pq}(t)$ by exchanging $q$ and $p$. {To
derive the coefficients (\ref{driftterm}) and (\ref{diffusionterm}) we have
employed the ansatz Eq.~(\ref{Fokker--Planck equation}) for the
Fokker--Planck operator. Acting with yet undefined functions }$f${$_{n}$,}$B$%
{$_{nm}$ ($n,m=p,q$) on the r.h.s. of Eq.~(\ref{Fokker--Planck equation}),
and comparing with its time derivative, yields conditions for the functions }%
$f${$_{n},B_{nm}$. An exact Fokker--Planck equation in the form of Eq.~(\ref%
{Fokker--Planck equation}) for the }${q}${--oscillator has been first
derived in Ref.~\cite{haak85} and later by a different method in Refs.~\cite%
{unr89,hu92}.}

\subsection{Purity}
\label{Purity}{\ }

A convenient quantity to measure the degree of global decoherence is the purity $\mathcal{P}(t)$,
defined as the average of the density matrix itself
\cite{zure93},%
\begin{equation}
\mathcal{P}(t)\equiv \langle \rho (t)\rangle \ =\ \tr\rho ^{2}(t)\ ,
\label{linear
entropy}
\end{equation}%
which is basis independent. Of special interest is the equilibrium value $%
\mathcal{P}_{\beta }\equiv \lim_{t\rightarrow \infty }\mathcal{P}(t)$ which
measures the efficiency of the environment in destroying quantum coherence.
Here we implicitly assume ergodic behavior which, as we shall see, applies
in the presence of Ohmic baths. For a harmonic oscillator in thermal
equilibrium, the reduced Wigner function is \cite{weis99}
\begin{equation}
W_{\beta }(q,p)=\frac{1}{2\pi \left[ \langle q^{2}\rangle _{\beta }\langle
p^{2}\rangle _{\beta }\right] ^{1/2}}\exp \left( -\frac{q^{2}}{2\langle
q^{2}\rangle _{\beta }}-\frac{p^{2}}{2\langle p^{2}\rangle _{\beta }}\right)
,
\end{equation}%
which leads to
\begin{equation}
\mathcal{P}_{\beta }=\frac{\hbar }{2\left[ \langle q^{2}\rangle _{\beta
}\langle p^{2}\rangle _{\beta }\right] ^{1/2}}\ .
\label{purityversuncertainty}
\end{equation}%

\section{Coherence decay for Ohmic damping}

\label{timedep} Although the equilibrium decoherence (as measured by the
product $\langle q^{2}\rangle _{\beta }\langle p^{2}\rangle _{\beta }$) is
enhanced by the additional noise term, one may wonder whether for low
temperatures the decoherence time becomes larger than for the damped $q$%
--oscillator. To answer this question exhaustively one would have to calculate the time evolution
of the purity for an arbitrarily pure initial condition. We consider two cases. First, we choose a
coherent (Gaussian) state as initial state. This case should present the greatest robustness
against decoherence \cite{zure93,kohl02}. Second, we choose a superposition of two Gaussian wave
packets. Then a new aspect of decoherence comes into play, namely, the fast vanishing of the
relative coherence between the two Gaussian wave packets. We distinguish the two manifestations of
decoherence by introducing a new quantity, which we call {\em relative purity}.

\subsection{Decoherence for a Gaussian wave packet as initial state}

\label{coherent state} We consider here the case where the system starts in
a coherent state at $t=0$. The Wigner function for a Gaussian wave packet is
\begin{equation}
W_{S}(q,p)=\frac{1}{\pi }\exp \left[ -\eta \left( q-q_{0}\right) ^{2}-\frac{1%
}{\eta }\left( p-p_{0}\right) ^{2}\right] \ ,
\label{wignerfuction of coherent state}
\end{equation}%
where $p_{0}=\sqrt{2\eta }\mathrm{Im\,}(\alpha )$ and $q_{0}=\sqrt{2/\eta }%
\mathrm{Re\,}(\alpha )$ are defined in terms of the complex eigenvalue $%
\alpha $ (amplitude) of the coherent state (recall $\eta =\sqrt{\omega
_{q}/\omega _{p}}$). Since $W_{S}(q,p,0)$ is Gaussian, the convolution
integral of Eq.~(\ref{wigsolution}) can be performed. The integrals become
particularly simple when both baths are equal ($\gamma _{p}=\gamma _{q}$ $%
=\gamma $ and $\eta =1$), corresponding to the completely \textit{symmetric
case}. Then we have $\phi _{p}=\phi _{q}$ $\equiv \phi _{1}$ and, by the
same token, $X_{pq}=X_{qp}\equiv X_{1}$ in Eq.~(\ref{definition of auxiliary
functions phi_0,phi_p}) and also $X_{p0}=X_{q0}$ $\equiv X_{0}$ in Eq.~(\ref%
{tempdep}). The crossed terms in Eq.~(\ref{wigsolution}) vanish and the
Wigner function becomes a product of two Gaussians for all times
\begin{eqnarray}
\fl\qquad W(q,p,t) &=&\frac{1}{4\pi }\frac{1}{\phi _{0}^{2}+\phi
_{1}^{2}+4X_{0}+4X_{1}}  \notag \\
&&\exp \left[ -\frac{(p-p_{0}\phi _{0}+q_{0}\phi _{1})^{2}+(q-q_{0}\phi
_{0}+p_{0}\phi _{1})^{2}}{\phi _{0}^{2}+\phi _{1}^{2}+4X_{1}+4X_{0}}\right] .
\label{wigner2}
\end{eqnarray}%
For the purity one obtains the simple expression
\begin{equation}
\mathcal{P}(t)\ =\ \frac{1}{\phi _{0}^{2}+\phi _{1}^{2}+4X_{0}+4X_{1}}\ .
\label{purity2}
\end{equation}%
The remaining task is to calculate the quantities $\phi _{0}$, $\phi _{1}$, $%
X_{0}$, $X_{1}$ from their definition in Eqs.~(\ref{definition of auxiliary
functions phi_0,phi_p}) and (\ref{tempdep}). Using the na\"{\i}ve form ($%
\Omega \rightarrow \infty $) of the spectral function of an Ohmic bath as
given in Sec.~\ref{Two independent baths}, we would find $\phi _{0}(0)$ $%
=1/(1+\gamma _{p}\gamma _{q})$ $\neq 1$ leading to inconsistencies [see the
discussion after Eq. \eref{tempdep}]. The reason for this lies in an initial
slippage caused by the somewhat unphysical character of the decoupled
initial condition \cite{legg87,bez80, hae97, sanc94}. Technically it stems
from the fact that for $J(\omega )$ $\propto \omega $ the integrals in Eqs.~(%
\ref{definition of auxiliary functions phi_0,phi_p}) do not converge at $t=0$
\cite{haak85}. This problem is overcome by regularizing the Ohmic spectral
functions, i.e. by reintroducing a finite cutoff $\Omega $. The explicit
calculation with a Drude regularized spectral function is sketched in {\ref%
{appB}}.\textbf{\ }Here we state only the main result: At $T=0$ the purity decays on two time
scales, given by $\Omega ^{-1}$ and $\tau$ in (\ref{tauspez}),
\begin{equation}
\mathcal{P}(t)\ \simeq \left\{
\begin{array}{cc}
e^{-\Omega t}~, & \text{for ~}0\leq t\lesssim \Omega ^{-1} \\
\left[\mathcal{\mathcal{P}}_{\beta } +\frac{1}{\omega_0^2t^{2}}\frac{2\gamma }
{1+\gamma^{2}}\cos(\Lambda t)e^{-t/\tau }\right] ,
 & \quad \ \text{for ~} \Omega ^{-1}\ll t\rightarrow \infty ~,%
\end{array}%
\right.   \label{purityresult}
\end{equation}%
with $\mathcal{\mathcal{P}}_{\beta }^{-1}=2\langle q^{2}\rangle _{\beta }$ and the oscillator frequency
$\Lambda =\omega_0/(1+\gamma^2)$. Coherence is reduced immediately after the start of the coupling. Although
afterwards it decreases more slowly, on a time scale $\tau $, for larger
couplings the curves of $\mathcal{P}(t)$ for different values of $\gamma $
never cross, i.e., $\mathcal{P}(t)$ is a monotonously decreasing function of
$\gamma $ for all $t$.

An initial slip similar to that discussed in this section also occurs for
the $q$--oscillator \cite{haak85}. However in that case its effect on purity
is much less severe. Specifically, out of the set of functions $\phi _{q}(t)$%
, $\phi _{p}(t)$ and $\phi _{0}(t)$ defined in Eq.~(\ref{definition of
auxiliary functions phi_0,phi_p}), only $\phi _{q}(t)$ is affected by the
initial slip. After a time $\Omega ^{-1}$ it becomes $\phi _{q}(t)\sim
\gamma _{p}$ $\neq 0$ \cite{haak85}. However, we note that $\phi _{q}$
appears in the temperature-dependent terms $X_{qp}$, $X_{pq}$, $Y_{qp}$, $%
Y_{pq}$ of Eq.~(\ref{tempdep}) only in combination with the spectral density
of the momentum coupling $J_{p}(\omega )$, which vanishes for the $q$%
--oscillator by definition. We thus reach the important conclusion that the
purity evolution of the $q$--oscillator is insensitive to the initial slip
stemming from the use of decoupled initial conditions:\ At $t\sim \Omega
^{-1}$ the purity is still approximately unity, decreasing afterwards at a
rate $\propto \gamma _{q}$.

Isar and coworkers \cite{isa99} studied in detail the decay of purity for the $q$--oscillator. In
particular they found constraints that must be satisfied by the bath if the purity is to remain
constant and close to unity during the whole time evolution of the oscillator. Such a high purity
is not possible in the present case due to the initial slip [see Eq.~(\ref{purityresult}].

Another possible preparation of the initial state is in a constrained equilibrium. The expectation
values $\langle q(0)\rangle$ $=$ $q_0$ and $\langle p(0)\rangle$ $=$ $p_0$ are held fixed with the
bath equilibrated around those values. Then the Wigner function is \cite{haak85}
\begin{equation}
W(q,p,t) = \frac{1}{2\pi \left[ \langle q^{2}\rangle _{\beta }\langle p^{2}\rangle _{\beta }\right]
^{1/2}}\exp \left( -\frac{(q-\langle q(t)\rangle)^{2}}{2\langle q^{2}\rangle _{\beta
}}-\frac{(p-\langle p(t)\rangle)^{2}}{2\langle p^{2}\rangle _{\beta }}\right)
\end{equation}
while the purity is given by its equilibrium value Eq.~(\ref{purityversuncertainty}). A more
detailed discussion of the purity decay of the $q$--oscillator, together with the general formulae,
is given in \ref{appB}.

\subsection{Decoherence of two Gaussian wave packets}

\label{towgaussian}

Formula (\ref{wigsolution}) also allows us to investigate more complicated
initial conditions such as as, for example, the superposition of two
Gaussian wave packets. This case has been studied for a single bath by
Caldeira and Leggett \cite{cald84}. It is an interesting case study because
it displays two different aspects of decoherence. On the one hand, there is
the decoherence which either wave packet would experience alone. This part
is essentially described by $\mathcal{P}(t)$, which we will call \emph{%
Gaussian purity} [see Eqs.~(\ref{purity2}) and (\ref{purityresult})]. On the
other hand there is the decoherence due to the spatial separation of the two
packets. This second contribution is expected to become increasingly
important when the distance $a$ between the two packets becomes large. As
initial wave function we choose
\begin{equation}
\psi (x)=\frac{1}{c\sqrt[4]{2\pi \sigma ^{2}}}\left( e^{-\frac{(x+a/2)^{2}}{%
4\sigma ^{2}}}+e^{-\frac{(x-a/2)^{2}}{4\sigma ^{2}}}\right) \ ,
\end{equation}%
which translates into an initial Wigner function
\begin{equation}
W_{S}(q,p)=\frac{1}{\pi  c^{2}}e^{-\frac{q^{2}}{2\sigma ^{2}}-2\sigma^{2}p^{2}}
\sum_{k=\pm 1}\left( e^{-\frac{a^{2}}{8}-%
\frac{kaq}{2\sigma ^{2}}}+e^{-ikap}\right) \ .  \label{wig2}
\end{equation}%
Here $c\equiv \sqrt{2}\left[ 1+\exp (-a^{2}/8\sigma ^{2})\right] ^{1/2}$ is
a normalization constant. We assume the most symmetric case and set $\sigma
=1/\sqrt{2}$. Plugging (\ref{wig2}) into (\ref{wigsolution}) we find that
the purity can be expressed in terms of the Gaussian purity as
\begin{equation}
P(t)=\frac{\mathcal{P}(t)}{2}\left\{ 1+\frac{\cosh ^{2}\left[ \frac{a^{2}}{4}%
(\phi (t)\mathcal{P}(t)-\frac{1}{2})\right] }{\cosh ^{2}\left(
a^{2}/8\right) }\right\} .  \label{two-purity}
\end{equation}%
The function $\phi (t)$ is given by
\begin{equation}
\phi (t)\ =\ \phi _{0}^{2}(t)+\phi _{1}^{2}(t)\ ,  \label{phi}
\end{equation}%
where $\phi _{0}$ and $\phi _{1}$ are defined in (\ref{definition of
auxiliary functions phi_0,phi_p}) and computed in (\ref{phifunc}) for the
symmetric case. $\phi (t)$ evolves from $\phi (0)=1$ to $\lim_{t\rightarrow
\infty }\phi (t)=0$. We define the \emph{relative purity} as the ratio
\begin{equation}
P_{\mathrm{rel}}(t)\ =\ P(t)/\mathcal{P}(t)  \label{prel}
\end{equation}%
As expected, $P_{\mathrm{rel}}(t)\rightarrow 1$ as $a\rightarrow 0$, and $P_{%
\mathrm{rel}}(t)\rightarrow 1/2$ as $a\rightarrow \infty $.

Interestingly, the structure of (\ref{two-purity}) is such that, as time
passes and $\phi (t)\mathcal{P}(t)$ evolves from $1$ to $\ 0$, the ratio $P_{%
\mathrm{rel}}(t)$ starts at unity, as corresponds to a pure state, then
decreases and finally, at long times, goes back to unity. When $a$ is large $%
P_{\mathrm{rel}}(t)$ decays rapidly to $1/2$, on a timescale $\sim
1/4a^{2}\gamma $. There it stays for a time which increases with distance as
$\sim \gamma ^{-1}\ln a$. Afterwards it returns to one. %When $a$ is large,
%$P_{\rm rel}(t)\simeq1/2$ most of the time except for $\phi\mathcal{P}$
%very close to 0 or 1.
The ratio 1/2 can be rightly interpreted as resulting from the incoherent
mixture of the two wave packets. Thus it comes as a relative surprise that $%
P_{\mathrm{rel}}(t)$ becomes unity again at long times, as if coherence
among the two wave packets were eventually recovered. The physical
explanation lies in the ergodic character of the long time evolution, with
both wave packets evolving towards the equilibrium configuration [see Eq.~(%
\ref{meanq}) and the subsequent discussion].

\section{Single bath with linear coupling to both position and momentum}

\label{onebath} It is instructive to compare Eq.~(\ref{mod5}) with the
Hamiltonian of a harmonic oscillator interacting with a single heat bath in
the most general form of linear coupling,
\begin{equation}
H=\frac{\omega _{q}}{2}q^{2}+\frac{\omega _{p}}{2}p^{2}+\sum \omega
_{k}\left\vert a_{k}+\frac{\lambda _{k}}{\omega _{k}}q+\frac{\mu _{k}}{%
\omega _{k}}p\right\vert ^{2}\ ,  \label{onebathonlyHamiltonian}
\end{equation}%
with complex parameters $\lambda _{k},\mu _{k}$. This model is different
from that discussed in the previous sections in that here time reversal
invariance is broken. By this we mean the following. In the models described
by Eq.~(\ref{mod5}), the bath modes might describe for
example a magnetic field coupled to the momentum of a charged particle,
which clearly would break time reversal invariance. However in that case,
such a symmetry breaking is somewhat fictitious since, due to the linear
nature of the coupling and to the modelling of the bath as a set of harmonic
oscillators, one can always find a unitary transformation which restores
time reversal invariance, i.e. which renders all parameters in the
Hamiltonian real quantities. It is easy to see that, for general (complex) $%
\lambda _{k}$ and $\mu _{k}$, such a unitary transformation cannot be found
for Eq.~(\ref{onebathonlyHamiltonian}).

By an analysis similar to that of \Sref{Two independent baths} one finds
general expressions for the symmetrized equilibrium correlation functions:
\begin{eqnarray}
\fl\quad C_{qq}^{(+)}(t) &=&\frac{1}{\pi }\int_{0}^{\infty }d\omega |\chi
(\omega )|^{2}\cos (\omega t)\coth (\hbar \beta \omega /2)\ \left\{ \mathrm{%
Im}\widetilde{J}_{q}(\omega )\left[ \omega _{p}+\mathrm{Re\,}\widetilde{J}%
_{p}(\omega )\right] ^{2}\right.  \notag
\label{symmetrized correlation function for one bath} \\
&&\lo\quad +\mathrm{Im}\widetilde{J}_{p}(\omega )\left[ \left( \omega +%
\mathrm{Re\,}\widehat{J}_{-}(\omega )\right) ^{2}+\left( \mathrm{Re\,}%
\widetilde{J}_{+}(\omega )\right) ^{2}\right]  \notag \\
&&\lo\quad -\left. \left[ \omega _{p}+\mathrm{Re\,}\widetilde{J}_{p}(\omega )%
\right] \mathrm{Im}\left[ 2\omega \widehat{J}_{-}(\omega )-\widehat{J}%
_{-}^{2}(\omega )-\widetilde{J}_{+}^{2}(\omega )\right] \right\} \ .
\end{eqnarray}%
The same expression applies for $C_{pp}^{(+)}(t)$ with the indexes $p,q$
interchanged everywhere. As before, we define a generalized
\textquotedblleft susceptibility" for the system which now reads
\begin{equation}
\fl\quad \chi ^{-1}(\omega )=\left[ \omega _{q}-\widetilde{J}_{q}(\omega )%
\right] \left[ \omega _{p}-\widetilde{J}_{p}(\omega )\right] -\omega ^{2}-%
\widetilde{J}_{+}^{2}(\omega )-\widehat{J}_{-}^{2}(\omega )+2\omega \widehat{%
J}_{-}(\omega )~.  \label{generalized susceptibility for one bath}
\end{equation}%
Eqs.~(\ref{symmetrized correlation function for one bath}) and (\ref%
{generalized susceptibility for one bath}) involves four different spectral
functions. $\widetilde{J}_{q}(\omega )$ and $\widetilde{J}_{p}(\omega )$ are
defined as in Eqs.~(\ref{mod6}) and (\ref{tildetransformed}). Now we
introduce the spectral functions
\begin{eqnarray}
J_{+}(\omega ) &=&\ \sum (\lambda _{k}\mu _{k}^{\ast }+\lambda _{k}^{\ast
}\mu _{k})\delta (\omega -\omega _{k})  \notag
\label{definitions of J_sym and J_asy} \\
&=&2\sum |\lambda _{k}||\mu _{k}|\cos \theta _{k}\delta (\omega -\omega _{k})
\notag \\
J_{-}(\omega ) &=&i\sum (\lambda _{k}\mu _{k}^{\ast }-\lambda _{k}^{\ast
}\mu _{k})\delta (\omega -\omega _{k})  \notag \\
&=&2\sum |\lambda _{k}||\mu _{k}|\sin \theta _{k}\delta (\omega -\omega
_{k})\ ,
\end{eqnarray}%
which reflect the mixing of the bath modes. The \textquotedblleft
hat\textquotedblright\ symbol denotes the transformation
\begin{equation}
\widehat{f}(\omega )\ \equiv \ \omega \mathcal{P}\int_{0}^{\infty }\frac{%
f(\omega ^{\prime })}{{\omega ^{\prime }}^{2}-\omega ^{2}}d\omega ^{\prime
}+i\frac{\pi }{2}f(|\omega |)\quad ,  \label{hattransformation}
\end{equation}%
In Eq.~(\ref{hattransformation}) the real part is an antisymmetric function
and the imaginary part is symmetric. This is exactly reverse to the
\textquotedblleft tilde" (Riemann) transform, defined in Eq.~(\ref%
{tildetransformed}).

Before we calculate Eq.~(\ref{symmetrized correlation function for one bath}%
) for a specific case, we analyze some generic features. First we notice
that, by setting $\widehat{J}_{-}(\omega )$ and $\widetilde{J}_{+}(\omega )$
to zero, we recover the autocorrelation function for two independent baths
[see Eq.~(\ref{symmetrized autocorrelationfkt})]. On the other hand, the two
spectral densities $J_{+}(\omega )$ and $J_{-}(\omega )$ are not independent
from $J_{p}(\omega )$ and $J_{q}(\omega )$; rather, they satisfy%
\begin{equation}
J_{+}^{2}(\omega )+J_{-}^{2}(\omega )=J_{p}(\omega )J_{q}(\omega )\ .
\label{relation between J_q,J_p and J_{sym},J_{asy}}
\end{equation}%
Actually this relation has already been used to simplify the integrand of
Eq.~(\ref{symmetrized correlation function for one bath}). We note that it
holds for the spectral densities themselves but in general \emph{not} for
their Riemann transforms $\widetilde{J}_{+}(\omega )$ and $\widehat{J}%
_{-}(\omega ) $. Apart from the above constraint one can, at least in
principle, freely choose three of the four spectral functions. If, in
addition, we make the physically reasonable assumption that the
\textquotedblleft mixing angle" in Eq.~(\ref{definitions of J_sym and J_asy}%
) is frequency independent ($\theta _{k}=\theta $), condition (\ref{relation
between J_q,J_p and J_{sym},J_{asy}}) fixes $J_{+}(\omega )$ and $%
J_{-}(\omega )$ completely.
\begin{eqnarray}  \label{symandasym}
J_{+}(\omega ) &=&\left[ J_{q}(\omega )J_{p}(\omega )\right] ^{1/2}\cos
\theta  \notag  \label{J_{sym} and J_{asy} for fixed eta} \\
J_{-}(\omega ) &=&\left[ J_{q}(\omega )J_{p}(\omega )\right] ^{1/2}\sin
\theta \ .
\end{eqnarray}%
Therefore, in the important case where both $J_{q}(\omega )$ and $%
J_{p}(\omega )$ have the same spectral exponent ($\alpha _{q}=\alpha
_{p}=\alpha $), $J_{+}$ and $J_{-}$ also obey the same power law. Using Eq.~(%
\ref{relation between J_q,J_p and J_{sym},J_{asy}}) we observe that the term
$(\mathrm{Im\,}\widetilde{J}_{q}\mathrm{Im\,}\widetilde{J}_{p})$ implicit in
Eq.~(\ref{generalized susceptibility for one bath}) drops out. We recall
that this interference term has been responsible for the non trivial
\textquotedblleft phase--space" diagram Fig. ~\ref{fig1}. What is more, only
$\mathrm{Re\,}\widetilde{J}_{+}$ appears in Eqs.~(\ref{symmetrized
correlation function for one bath}) and (\ref{generalized susceptibility for
one bath}). That means, according to Eq.~(\ref{ReJn}), that for $\alpha
_{q}+\alpha _{p}<4$ the symmetric spectral function $J_{+}$ does not
contribute at all. The relation between the \textquotedblleft hat" and the
\textquotedblleft tilde" transformation is
\begin{equation}
\widehat{f}(\omega )=\widetilde{\omega f(\omega )}\ ,  \label{hattilde}
\end{equation}%
from which all properties of $\widehat{f}$ can easily be deduced. For
example from Eq. (\ref{hattilde}) and Eqs.~(\ref{ReJn}) and ~(\ref{gen1}) it
follows that $\mathrm{\mathrm{Re}}\widehat{J}_{-}(\omega )=0$ only for $%
\alpha _{q}+\alpha _{p}<2$.

It is illustrative to look at the \textquotedblleft
generalized\textquotedblright\ susceptibility Eq.~(\ref{generalized
susceptibility for one bath}) for the case $\alpha _{q}+\alpha _{p}<2$,
where all real parts of the spectral functions vanish. We obtain
\begin{eqnarray}
\fl\quad \chi ^{-1}(\omega ) &=&\omega _{0}^{2}-\omega ^{2}-\omega _{q}%
\widetilde{J}_{p}(\omega )-\omega _{p}\widetilde{J}_{q}(\omega )+2\omega
\widehat{J}_{-}(\omega )~  \notag \\
&=&\omega _{0}^{2}-\omega ^{2}-\mathrm{sgn\,}(\omega )\frac{i\pi \omega _{p}%
}{2}\left[ J_{q}(|\omega |)+\eta ^{2}J_{p}(|\omega |)-\frac{2|\omega |}{%
\omega _{p}}J_{-}(|\omega |)\right] \ .
\label{generalized susceptibility for one bath1}
\end{eqnarray}%
This is exactly the suceptibility of a $q$--oscillator coupled to a single
bath with an additive spectral function
\begin{equation}
\label{effsingle}
J(\omega )\ =\ J_{q}(\omega )+\eta ^{2}J_{p}(\omega )+\frac{2\omega
J_{-}(\omega )}{\omega _{p}}.
\end{equation}%
We conclude that, when only one bath is involved, the particular structure
of the coupling of the bath to the position or momentum variable can always
be modelled by a $q$--oscillator with an appropriately chosen spectral
function. In this context the combination $\omega J_{-}(\omega )$ can be
considered as stemming from an effective additional noise source that
couples to position. Note however, that this is not the same as saying that a 
single--bath coupling to $q$ and $p$ can always be unitarily transformed to a 
coupling only to $q$ ($q$--oscillator), since this is not true in general. For one thing, the noise 
properties, which are not uniquely given by $\chi(\omega)$, would be different.

We also wish to emphasize that \textit{a double-bath
dissipative oscillator cannot be described in terms of an oscillator coupled
to an effective single bath}. In particular, it can never be modelled by a
susceptibility like (\ref{effsingle}). This is the reason why the physics explored in %
\Sref{Two independent baths} is so different from that which could be found
in any possible single-bath scenario.

We do not repeat here the analysis of \Sref{Two independent baths} for
arbitrary spectral functions but focus instead on the case of Ohmic coupling
($\alpha _{q}=\alpha _{p}=1$) in order to compare with the results obtained
for two independent baths. As $\mathrm{Re}\widetilde{J}_{n}(\omega )=\mathrm{%
Re}\widetilde{J}_{+}(\omega )=0$ vanish [see \Sref{gencase}], we obtain
\begin{eqnarray}
&&C_{qq}^{(+)}(t)\ =\frac{1}{\pi }\int_{0}^{\infty }d\omega |\chi (\omega
)|^{2}\cos (\omega t)\coth (\hbar \beta \omega /2)  \notag
\label{ohmic symmetrized correlation} \\
\mathrm{Im} &&\left\{ \omega _{q}^{2}\widetilde{J}_{q}(\omega )+\widetilde{J}%
_{p}(\omega )\left[ \omega -\mathrm{Re\,}\widehat{J}_{-}(\omega )\right]
^{2}-2\omega _{p}\widehat{J}_{-}(\omega )\left[ \omega -\mathrm{Re\,}%
\widehat{J}_{-}(\omega )\right] \right\} .
\end{eqnarray}%
The form of the susceptibility varies depending on whether the mixing angle
is a multiple of $\pi $ or not. We discuss the two cases separately.

\subsection{Mixing angle $\protect\theta =0,\protect\pi $}

\label{eta=0} 
\begin{figure}[tbp]
\begin{center}
\epsfig{figure=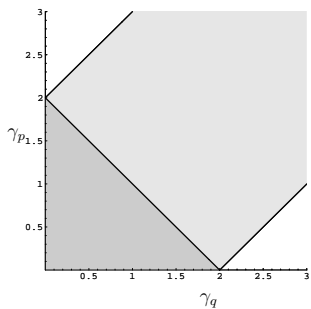,width=4cm,height=4cm}
\end{center}
\caption{ Regions of underdamped oscillations for a single bath coupling to
position and momentum (dark region) and for two independent baths coupling
to position and momentum (dark and bright region). $\protect\eta =1$.}
\label{fig3}
\end{figure}

For $\theta =0$ or $\theta =\pi $ one can bring, by an
appropriate redefinition of $a_{k}$, $a_{k}^{\dagger }$, the interaction
part of the Hamiltonian Eq.~(\ref{onebathonlyHamiltonian}) into the form
\begin{eqnarray}
H_{I}&\propto & \sum q_{k}(|\mu _{k}|p+|\lambda _{k}|q)\ ,
\label{forms of interaction}
\end{eqnarray}%
so that the bath couples to the main oscillator either through the position
of their oscillators or through their momentum, but not through both. % We
% observe that this interaction Hamiltonian is similar to the coupling in Eq.~(%
% \ref{mod5}) to each of the two independent baths.
From Eq.~(\ref{symandasym}) it can be seen that the antisymmetric spectral
function vanishes identically. Apart from the susceptibility, the integrand
of Eq.~(\ref{ohmic symmetrized correlation}) becomes equal to that of Eq.~(%
\ref{symmetrized autocorrelationfkt}). Thus all results of Sec.~\ref{ohm},
in particular Eqs.~(\ref{classical}) and (\ref{meanq}), carry formally over.
The only but crucial difference lies in the poles of the generalized
susceptibility $\chi (\omega )$. The set of zeros can be written as 
\begin{equation}
\label{zerosonebath}
\omega_{\pm} \ = \ \omega_0\left(-i\kappa/2\pm \sqrt{1-\kappa^2}\right)  \ ,
\end{equation}
%in Eq.~(%\ref{solu}), 
with $\kappa $ defined as
\begin{equation}
\kappa =\frac{\eta }{2}\gamma _{p}+\frac{1}{2\eta }\gamma _{q}\ .
\label{kapp1}
\end{equation}%
According to criterion A, Eq.~(\ref{crit1}), the regions of overdamped and
underdamped oscillations in the $(\gamma _{q},\gamma _{p})$ plane are
separated by a straight line with negative slope given as before by the
condition $\kappa =1$ [see Fig.~\ref{fig3}].
Therefore, a crossover from overdamped to underdamped oscillations by
increasing one of the coupling constants is now impossible. We also note that, 
in the symmetric case ($\gamma_q$ $=\gamma_p$ $=\gamma$), 
the relaxation time is the same as for the $q$--oscillator 
[$\tau^{-1}$ $=\omega_0\gamma/2$]. Increasing $\gamma _{q}$ or $\gamma _{p}$ leads 
inevitably to an enhancement of
dissipation. This is in blatant \textit{contrast} to the results obtained
previously for the double-bath model and one of the most remarkable results
of this work. It best illustrates the importance of the specific structure
of the bath and justifies a posteriori the detailed study of a model with
two independent baths.

\subsection{Mixing angle $\protect\theta \neq 0,\protect\pi $}

We briefly consider the case of non--vanishing mixing angle. The effects
should be largest for $\theta =\pi /2$, corresponding to an interaction
Hamiltonian
\begin{equation}
H_{I}\propto \sum (|\mu _{k}|p_{k}p+|\lambda _{k}|q_{k}q)\ .
\end{equation}%
This form of coupling is called amplitude coupling in \cite{gor04}.
Now $J_{-}(\omega )$ is not longer zero but we have
\begin{equation}
\omega \widehat{J}_{-}(\omega )=\sqrt{\gamma _{q}\gamma _{p}}\omega ^{2}%
\left[ \frac{2}{\pi }\ln \left( 1+\frac{\Omega ^{2}}{\omega ^{2}}\right) +i%
\mathrm{sgn\,}(\omega )\right]  \label{ohmicJtilde}
\end{equation}%
which is exactly the spectral function of a superohmic bath with exponent $%
\alpha =2$. The logarithm in $\mathrm{Re\,}\widehat{J}_{-}$ inhibits further
analytic treatment of Eq.~(\ref{ohmic symmetrized correlation}) even in the
high temperature limit; thus we limit ourselves to a qualitative analysis.
For simplicity, we omit the terms containing logarithms in the generalized
suceptibility Eq.~(\ref{generalized susceptibility for one bath}). The
contributions to $C_{qq}^{(+)}$ stemming from those terms should decay on a
time scale $\sim \Omega ^{-1}$. The remaining terms define a suceptibility
which is identical to the suceptibility of a $q$--oscillator coupled
linearly to one bath with spectral density%
\begin{equation}
J(\omega )=(\gamma _{q}+\eta^2 \gamma _{p})\omega +\frac{2\sqrt{\gamma _{q}\gamma _{p}}}
           {\omega_{p}}\,\omega ^{2}~.
\end{equation}%
It is well known \cite{weis99} that, in the case of a polynomial spectral
function, the term with the lowest exponent dominates [see also the
discussion in Sec.~\ref{gencase}]. Therefore we may conclude that the
crossover diagram \Fref{fig3} remains unchanged by a mixing angle $\theta
\neq 0$.

\section{Conclusions}

We have discussed the behavior of a quantum Brownian particle in a harmonic
potential subject to two independent noise sources, one of which couples to
its position and the other one to its momentum. In the symmetric case where
both baths are Ohmic and their coupling strength is the same, we find
underdamped oscillations of the central oscillator for all coupling
strengths. This indicates that the two baths partially cancel each other.
The effect is due to the mutually conjugate character of position and
momentum. It was first noted in Ref.~\cite{cast03} for the (cylindrically)
symmetric spin--boson model with spin $s=\frac{1}{2}$, i.e. in the deep quantum
regime. \textquotedblleft Quantum frustration" of the spin can pictorially
be described as the result of having two observers attempting to measure
simultaneously, with equal efficiency, two non-commuting components of the
spin. Because of the uncertainty principle, both of them fail to measure
anything. Here we have investigated the analogous effect for a quantum
oscillator coupled to two independent baths. We have found a moderate form
of cancellation which we have labelled \textquotedblleft quasiclassical
frustration" because our dissipative oscillator may describe a large spin
impurity coupled to the magnon bath in a ferromagnetic medium. 
The most notable features of quasiclassical frustration become manifest in 
the strong coupling limit, where underdamped dynamics survives and all 
time scales diverge. It remains to
be investigated whether the occurrence of frustration in the classical
regime is a general property or an artifact of the harmonic oscillator.

%The situation is reminiscent of the so-called quantum Zeno effect. There an
%observer measures an observable in its eigenstate and thereby inhibits or at
%least slows down the quantum transition into other states \cite%
%{chi77,ita90,boc91}. If one considers the central oscillator plus one bath
%as the system and the other bath as the measurer, the system
%considered here shares features such
%as the dilatation of all time scales and the freezing of the dynamics (see
%Sec.~\ref{ohm}) with the quantum Zeno effect. However one has to be most careful in interpreting
%the observed effects as a Quantum Zeno effect
%due to the classical nature of the harmonic oscillator.

We have compared the double-bath model with the case where a single bath
couples linearly to the position and momentum of the oscillator. In the
latter case the \textquotedblleft phase--space\textquotedblright\ diagram is
simpler in that transitions from overdamped to underdamped oscillations can
never occur. Comparison of the two models indicates that bath correlations
can qualitatively change the behaviour of a dissipative system.

A point of caution is needed in the interpretation of our results. We have
definitely ruled out the at first sight enticing but at closer inspection
unphysical conjecture that the effects of the two observers cancel each
other completely and the particle is not affected by the environment. Indeed
we have seen in the specific example of a decoupled initial state that, in
destroying quantum coherence, two baths are always more efficient than one.
This is true for both the overdamped and the underdamped regime. This
demonstrates that, at least for the harmonic oscillator, underdamped
dynamics is by no means a reliable signature of high global coherence, which
here we identify with purity [see Sec. \ref{timedep}].
% More work is needed
% to decide whether this is a general feature of dissipative quantum systems
% or just an artifact of the harmonic oscillator .
Our work provides further evidence that decoherence and dissipation are not
necessarily correlated. By dissipation we mean here the net transfer of
energy from the central oscillator to the thermal baths. It is characterized
by the classical equation of motion or, equivalently, by the properties of
the spectral function. We have seen that, depending on the situation, \ an
increase of dissipation can be accompanied by either a reduction or an
increase of decoherence. Vice versa, a source of decoherence may or may not
lead to dissipation. This is the case in the so called pure dephasing (not
considered here), where the interaction part of the Hamiltonian commutes
with the system Hamiltonian and no dissipation occurs at all, see i.~e.~\cite%
{gor04}.

The fact that we have focused on a quantum oscillator coupled linearly to
oscillator baths has allowed us to investigate analytically the equilibrium
and dynamical properties. The question of frustration is also raised in
other dissipative quantum systems, such as a small spin coupled to a boson
\cite{cast03} or a fermion \cite{sols87} bath, which are not amenable to an
exact analytical treatment. The extension of the present study to less
tractable physical scenarios provides a theoretical challenge.

\ack We are indebted to F.~Guinea for useful discussions. This work has been
supported by Ministerio de Ciencia y Tecnolog\'{\i}a (Spain) under Grants
No. BFM2001-0172 and FIS2004-05120, and by the Ram\'{o}n Areces Foundation.
One of us (H.K.) acknowledges financial support from the RTN Network of the
European Union under Grant No.~HPRN--CT--2000-00144.

\appendix

\section{Calculation of Eq.~(\protect\ref{highcont})}

\label{appA} We calculate the high frequency contribution $\mathcal{C}%
_{\Omega }$ to the position mean square in Eq.~(\ref{expression for general
spectralfunctions}). This contribution vanishes for $\alpha _{q}+\alpha
_{p}<2$ but yields a finite contribution otherwise. We assume again $\alpha
_{q}=\beta _{q}/m$ and $\alpha _{p}=\beta _{p}/m$, ($\beta _{q}$,$\beta _{p}$%
,$m$ $\in \mathbb{Z}$) to be rational numbers. Then this contribution can be
written in a determinant form %\begin{equation} \label{cutoff contibution}
%\mathcal{C}_{\Omega }=\frac{1}{\gamma _{p}\gamma
%_{q}^{2}}\frac{m}{\pi \Delta _{r_{\Omega }}(\lambda ^{2})}\left\vert
%\begin{array}{cccc}
%f(\lambda _{1}^{2}) & \lambda _{1}^{2r-4} & \cdots  & 1 \\
%\vdots  & \vdots  & \vdots  & \vdots
%\end{array}%
%\right\vert \ .
%\end{equation}%

%\begin{equation}
%\mathcal{C}_{\Omega }=\frac{1}{\gamma _{p}\gamma _{q}^{2}}\frac{m}{\pi
%\Delta _{r_{\Omega }}(\lambda ^{2})}\left\vert
%\begin{array}{cccc}
%f(\lambda _{1}^{2}) & \lambda _{1}^{2r-4} & \ldots  & 1\cr\vdots  & \vdots
%& \vdots
%\end{array}%
%\right\vert \ .  \notag  \label{cutoff contibution} \\
%\end{equation}%

\begin{equation}
\mathcal{C}_{\Omega }=\frac{1}{\gamma _{p}\gamma _{q}^{2}}\frac{m}{\pi
\Delta _{r_{\Omega }}(\lambda ^{2})}\left\vert
\begin{array}{cccc}
f(\lambda _{1}^{2}) & \lambda _{1}^{2r-4} & \ldots  & 1 \\
\vdots  & \vdots  & \vdots  & \vdots
\end{array}%
\right\vert   \label{cutoff contibution}
\end{equation}%
Here we use the Vandermonde determinant $\Delta _{N}(\lambda
^{2})=\prod_{i<j}^{N}(\lambda _{i}^{2}-\lambda _{j}^{2})$. We recall that $r$
is the total number of poles, while $r_{\Omega }<r$ is the number of high
energy poles. The function $f(x)$ is given by
\begin{eqnarray}
f(x) &=&\Phi (x,\beta _{p}+3m-1,r)+\gamma _{q}\gamma _{p}\frac{4\Theta
(\alpha _{p}-2)}{\pi ^{2}(\alpha _{p}-2)^{2}}\Phi (x,5m+\beta _{q}-1,r)
\notag  \label{ffunction} \\
&&\qquad \qquad +\gamma _{q}\gamma _{p}\Phi (x,2\beta _{p}+\beta
_{q}+m-1,r)\ ,
\end{eqnarray}%
where the function $\Phi (x,s,r)$ is a rather complicated function involving
a special function called Lerch transcendent \cite{gra00}. It can, for $s$
odd, be expressed in terms of a logarithm plus a polynomial
\begin{equation}
\Phi (x,s,r)=\frac{x^{s-1}}{2}\left[ \ln \left( 1+\frac{1}{x^{2}}\right)
-\sum_{n=1}^{(s+1)/2-r}\frac{(-1)^{n}}{nx^{2n}}\right] \ .
\label{phifunction}
\end{equation}%
In order to use this formula we assume $m$ and $\beta _{n}$ odd.
Consequently we formally exclude $\alpha _{n}$ even. However, that case can
be included by considering the limit $m\rightarrow \infty $ with $\beta _{n}$
$=2nm-1$. The argument of $f(x)$ is the square of one of the $r$ roots of
the characteristic polynomial
\begin{equation}
\fl\qquad \left[ \frac{2\Theta (\alpha _{q}-2)\lambda ^{m}}{\pi (\alpha
_{q}-2)}-i\lambda ^{\beta _{q}-m}\right] \left[ \frac{2\Theta (\alpha
_{p}-2)\lambda ^{m}}{\pi (\alpha _{p}-2)}-i\lambda ^{\beta _{p}-m}\right] -%
\frac{1}{\gamma _{q}\gamma _{p}}=0\ .  \label{highfrequcharacteristicpoly}
\end{equation}%
The asymptotic behavior of $\mathcal{C}_{\Omega }$ for $\gamma _{q},\gamma
_{p}\ll 1$ can be derived by expanding $\Phi $ in Eq.~(\ref{phifunction}).
For $\alpha _{p}>1$ we find to lowest order
\begin{equation}
\fl\qquad \mathcal{C}_{\Omega }=\frac{\gamma _{p}}{\pi }\frac{1}{\alpha
_{p}-1}+\gamma _{q}\gamma _{p}\frac{4\Theta (\alpha _{p}-2)}{\pi ^{2}(\alpha
_{p}-2)^{2}}\frac{1}{\alpha _{q}+1}+\frac{\gamma _{q}\gamma _{p}}{2\alpha
_{p}+\alpha _{q}-3}+\mathcal{O}(\gamma _{q}^{2},\gamma _{p}^{2})\ .
\label{boring}
\end{equation}%
Keeping only the first term for small $\gamma _{q}$ and small $\gamma _{p}$
yields Eq.~(\ref{highcont}).

\section{Derivation of Eq.~(\protect\ref{purityresult})}

\label{AppBa} We regularize the Ohmic spectral functions by a Drude cutoff
\begin{equation}
J_{n}(\omega )\ =\ \frac{2\gamma _{n}}{\pi }\frac{\Omega _{n}^{2}\omega }{%
\Omega _{n}^{2}+\omega ^{2}}\ ,\quad \widetilde{J}_{n}(\omega )\ =\ \frac{%
\gamma _{n}\omega \Omega _{n}}{\omega -i\Omega _{n}}\ ,\quad n=q,\ p\ .
\label{drude}
\end{equation}%
Assume $\Omega _{q}$ $=\Omega _{p}$ $=\Omega $, $\gamma _{q}=$ $\gamma
_{p}=\gamma $. For $\phi _{0}$ and $\phi _{1}$ we obtain
\begin{eqnarray}
\ \phi _{0}(t) &=&\frac{1}{1+\gamma ^{2}}\left[ \cos \left( \Lambda t\right)
-\gamma \sin \left( \Lambda t\right) \right] e^{-t/\tau }  \notag \\
&&\qquad \quad +\frac{\gamma ^{2}}{1+\gamma ^{2}}\left[ \cos (\gamma
^{2}\Omega t)+\frac{1}{\gamma }\sin (\gamma ^{2}\Omega t)\right] e^{-\Omega
t}  \notag \\
\phi _{1}(t) &=&\frac{1}{1+\gamma ^{2}}\left[ \gamma \cos \left( \Lambda
t\right) +\sin \left( \Lambda t\right) \right] e^{-t/\tau }  \notag \\
&&\qquad \quad +\frac{\gamma ^{2}}{1+\gamma ^{2}}\left[ -\cos (\gamma
^{2}\Omega t)+\gamma \sin (\gamma ^{2}\Omega t)\right] e^{-\Omega t}\ ,
\label{phifunc}
\end{eqnarray}%
where no ambiguity is left [$\Lambda = \omega _{0}(1+\gamma ^{2})^{-1}$
and, we recall, $\tau ^{-1}$ $=\gamma \Lambda $]. In a time of order $\Omega
^{-1}$ after the connection (with $\Omega \rightarrow \infty $), both
functions have dropped on average to a value $(1+\gamma ^{2})^{-1}$. This is
the value one would have obtained by using directly $J(\omega )$ $\propto
\omega $. After this short initial period, $\phi _{0}$ and $\phi _{1}$ decay
much more slowly, on a time scale $\tau $. In the
expression (\ref{purity2}) for the purity decay after the initial slip, we
may safely neglect the first two terms in the denominator. The functions $%
\phi _{0}$, $\phi _{1}$ however still govern through $X_{0}$, $X_{1}$ the
time evolution of the purity $\mathcal{P}(t)$ [see Eq. (\ref{tempdep})]. We
notice that both $X_{0}$ and $X_{1}$ are zero at $t=0$. However in the
initial time interval of order $1/\Omega $ they both increase rapidly. After
the initial slip, they settle to a value $\approx \langle q^{2}\rangle
_{\beta }\gamma ^{2}/$ $(1+\gamma ^{2})$ and increase much more slowly
afterwards. This can be seen if we write for instance $X_{0}(t)$ as follows
\begin{eqnarray}
X_{0}(t) &=&\frac{1}{8}\int d\omega J(\omega )\coth \left( \beta \omega
/2\right) |\chi (\omega )|^{2}\omega ^{2}\left\vert 1-g_{0}(\omega
,t)e^{-i\omega t}\right\vert ^{2}\ ,  \label{X0exp}
\end{eqnarray}%
where $g_{0}(\omega ,t)$ is the principal value integral
\begin{equation}
g_{0}(\omega ,t)\ =\ \frac{1}{2\pi i}\frac{1}{\chi (\omega )\omega}\mathcal{P}%
\int \frac{\chi (\omega ^{\prime })\omega ^{\prime }}{\omega ^{\prime
}-\omega }e^{i\omega ^{\prime }t}d\omega ^{\prime }\ .  \label{g0aux}
\end{equation}%
Eq.~(\ref{g0aux}) can be evaluated straightforwardly using the residue
theorem. However for our purposes the most important point is that $%
g_{0}(\omega ,0)$ $=1$ and that $g_{0}(\omega ,t)$ behaves in time
essentially in the same way as $\phi _{0}(t)$ in Eq.~(\ref{phifunc}). Thus
we may write
\begin{eqnarray}
\fl\ g_{0}(\omega ,t) &=&\frac{1}{\omega(1+\gamma ^{2})}\left[ \omega \cos \left( \Lambda t\right)
- i \left( \omega _{0}+\widetilde{J}(\omega)\right) \sin \left( \Lambda t\right) %
\right] e^{-t/\tau }  \notag \\
\fl &&\qquad \quad +\frac{\gamma ^{2}}{1+\gamma ^{2}}\left[ A_{\Omega
}(\omega )\cos (\gamma ^{2}\Omega t)+B_{\Omega }(\omega )\sin (\gamma
^{2}\Omega t)\right] e^{-\Omega t}\ .  \label{aux1}
\end{eqnarray}%
Here $A_{\Omega }$, $B_{\Omega }$ are some algebraic functions of $\omega $
whose exact form is not important for the present discussion, since after
the initial slip of time $\Omega ^{-1}$ the second line of Eq.~(\ref{aux1})
can be neglected. Proceeding in the same way with $X_{1}(t)$ we arrive at
\begin{equation}
\fl2X_{0}(t)+2X_{1}(t)\ =\ \langle q^{2}\rangle _{\beta }\left( 1+\frac{%
e^{-2t/\tau }}{(1+\gamma ^{2})^2}\right) +\frac{2C_{qq}^{(+)}(t)}{1+\gamma ^{2}}%
\cos \left( \Lambda t\right) e^{- t/\tau }\ ,  \label{endresult}
\end{equation}%
where $C_{qq}^{(+)}(t)$ is given in Eq.~(\ref{symmetrized autocorrelationfkt}%
). % The damping time is given by
% \begin{equation}
% \tau_{\rm dp}=\frac{\gamma\omega_0}{1+\gamma^2} \ .
% \label{damptime}
% \end{equation}
The temperature dependence is encoded in $\langle q^{2}\rangle _{\beta }$
and $C_{qq}^{(+)}(t)$ . Of course the r.h.s.~of Eq.~(\ref{endresult}) is not
zero any more at $t=0$ due to the neglected initial transient. At zero
temperature the autocorrelation function decays algebraically $\propto
-\omega_0^2 t^{-2} $. Then we reproduce Eq.~(\ref{purityresult}).

\section{Purity decay of the $q$--oscillator and general formulae}

\label{appB} Here we state the general formulae for the purity decay of a
coherent (Gaussian) initial state. To calculate the purity decay for two
Ohmic baths, both with decoupled initial conditions but with different
coupling strengths, we proceed as in Sec.~\ref{coherent state}. Performing a
twofold Fourier transform of Eq.~(\ref{wignerfuction of coherent state}) and
plugging the result into Eq.~(\ref{wigsolution}) yields for the reduced
Wigner function
\begin{equation}
W(q,p,t)=\frac{1}{\pi \left( A_{q}A_{p}-4A_{qp}^{2}\right) ^{1/2}}\exp
\left( \frac{4A_{qp}qp-A_{q}q^{2}-A_{p}p^{2}}{A_{q}A_{p}-4A_{qp}^{2}}\right)
\ .
\end{equation}%
Here,
\begin{eqnarray}
A_{q} &\equiv &\phi _{0}^{2}\eta +\frac{\phi _{q}^{2}}{\eta }+4X_{pq}+4X_{q0}
\notag \\
A_{p} &\equiv &\frac{\phi _{0}^{2}}{\eta }+\phi _{p}^{2}\eta +4X_{qp}+4X_{p0}
\notag \\
A_{qp} &\equiv &\frac{\eta \phi _{0}\phi _{p}}{2}-\frac{\phi _{0}\phi _{q}}{%
2\eta }-2Y_{pq}+2Y_{qp}\ .  \label{appB2}
\end{eqnarray}%
For the purity we have with the above definitions
\begin{equation}
\mathcal{P}(t)=\left( A_{q}A_{p}-4A_{qp}^{2}\right) ^{-1/2}\ .  \label{appB3}
\end{equation}%
As expected, this expression reduces to Eq.~(\ref{purity2}) for $J_{q}=J_{p}$
and $\eta =1$. For $J_{p}=0$ we have the case of the $q$--oscillator. From
the definitions~(\ref{tempdep}) we get $X_{pq}=X_{p0}$ $=Y_{pq}=0$ and
obtain a closed expression for the purity decay of the $q$--oscillator,
\begin{eqnarray}
\mathcal{P}(t) &=&\left[ \left( \phi _{0}^{2}\eta +\frac{\phi _{q}^{2}}{\eta
}+4X_{q0}\right) \left( \frac{\phi _{0}^{2}}{\eta }+\phi _{p}^{2}\eta
+4X_{qp}\right) -\right.  \notag  \label{appB4} \\
&&\qquad \qquad \left. 4\left( \frac{\eta \phi _{0}\phi _{p}}{2}-\frac{\phi
_{0}\phi _{q}}{2\eta }+2Y_{qp}\right) ^{2}\right] ^{-1/2}
\end{eqnarray}%
The explicit form of the functions $\phi _{0}(t)$, $\phi _{n}(t)$ with Drude
regularized Ohmic spectral function can be found in Ref. \cite{haak85}. Here
we give only their form after taking the limit $\Omega \rightarrow \infty $,
i.~e.~for $t\gg \Omega ^{-1}$.
\begin{eqnarray}
\phi _{0}(t) &=&\left[ \cos (\zeta t)-\frac{1}{\zeta \tau }\sin (\zeta t)%
\right] e^{-t/\tau }  \notag  \label{appB5} \\
\phi _{p}(t) &=&\frac{\omega _{p}}{\zeta }\sin (\zeta t)e^{-t/\tau }  \notag
\\
\phi _{q}(t) &=&-\gamma _{q}-\gamma _{q}\left[ \cos (\zeta t)-\frac{1}{\zeta
\tau }\sin (\zeta t)\right] e^{-t/\tau }\
\end{eqnarray}%
As in the main text $\zeta $ is the renormalized oscillator frequency. With
these functions we can calculate $X_{q0}(t)$, $X_{qp}(t)$ and $Y_{qp}(t)$.
We obtain,
\begin{eqnarray}
X_{q0}(t) &=&\frac{1}{2}\langle p^{2}\rangle _{\beta }\left[ 1+\phi
_{0}^{2}(t)\right] -\phi _{0}(t)C_{qq}^{(+)}(t)+\frac{\dot{\phi}_{0}^{2}(t)}{%
2\omega _{p}^{2}}\langle q^{2}\rangle _{\beta }+\frac{\dot{\phi}_{0}^{2}}{%
\omega _{p}^{2}}\dot{C}_{qq}^{(+)}(t)  \notag  \label{appB6} \\
X_{qp}(t) &=&\frac{1}{2}\langle q^{2}\rangle _{\beta }\left[ 1+\phi
_{0}^{2}(t)\right] -\phi _{0}(t)C_{qq}(t)+\frac{1}{2}\phi _{p}^{2}(t)\langle
p^{2}\rangle _{\beta }-\frac{\phi _{p}(t)}{\omega _{p}}\dot{C}_{qq}^{(+)}(t)
\notag \\
Y_{qp}(t) &=&\frac{\dot{\phi}_{0}(t)}{2\omega _{p}}\left[ \phi
_{0}(t)\langle q^{2}\rangle _{\beta }-C_{qq}^{(+)}(t)\right] +\frac{\phi _{p}%
}{2}\left[ \langle p^{2}\rangle _{\beta }\phi _{0}(t)-C_{pp}^{(+)}(t)\right]
\end{eqnarray}%
We notice that
\begin{equation}
Y_{qp}(0)\ =\ X_{q0}(0)\ =\ X_{qp}(0)\ =\phi _{p}(0)\ =\ 0\ .  \label{appB7}
\end{equation}%
Therefore we find from Eq.~(\ref{appB4}), $\mathcal{P}(0)=1$, i.e., the
purity is \textit{unaffected} by the initial slip.

%\end{references}

\end{document}